\documentclass[11pt,a4paper]{article}
\usepackage{jheppub}
 \usepackage{graphicx}
   \usepackage{amsmath,amsthm,amssymb}
     \usepackage{pifont}
      \usepackage{srcltx}

\newcommand{\nn}{\nonumber}

\newcommand{\tr}{\mathrm{tr}}

\renewcommand{\(}{\left(}
\renewcommand{\)}{\right)}
\renewcommand{\[}{\left[}
\renewcommand{\]}{\right]}

\title{Exponentiation for products of Wilson lines\\within the generating function approach}
\author{~A.A.Vladimirov}
\affiliation{ Department of Astronomy and Theoretical Physics, Lund University,\\ \small S\"olvegatan 14A, S 223 62 Lund, Sweden}
\emailAdd{vladimirov.aleksey@gmail.com}

\preprint{\begin{flushright} LU TP 15-02\\  January 2015 \end{flushright}}

\abstract{We present the generating function approach to the perturbative exponentiation of correlators of a product of Wilson lines and loops.
The exponentiated expression is presented in closed form as an algebraic function of correlators of known operators, which can be seen as a
generating function for web diagrams. The expression is naturally split onto two parts: the exponentiation kernel, which accumulates all
non-trivial information about web diagrams, and the defect of exponentiation, which reconstructs the matrix exponent and is a function of the
exponentiation kernel. The detailed comparison of the presented approach with existing approaches to exponentiation is presented as well. We
also give examples of calculations within the generating function exponentiation, namely, we consider different configurations of light-like
Wilson lines in the multi-gluon-exchange-webs (MGEW) approximation. Within this approximation the corresponding correlators can be calculated
exactly at any order of perturbative expansion by only algebraic manipulations. The MGEW approximation shows violation of the dipole formula for
infrared singularities at three-loop order.}
\begin{document}
 \maketitle

\section{Introduction}

The dynamics of Wilson lines and loops is responsible for nearly every aspect of gauge theories. Naturally, the correlators of Wilson lines and
loops are one of the most attractive objects of theoretical investigations. In this article we present the generating functional approach to the
exponentiation of a general correlator of Wilson lines.

The exponentiation of a correlator of Wilson lines is the fundamental property of the perturbative approach. As the name suggests, the
exponentiation property allows one to present the perturbative series for the correlator as an exponent. The non-triviality of this statement
consists in the special diagrammatic of the exponentiated expression. Namely, only connected diagrams enter the argument of the exponent.
Moreover, often the exponentiated diagrams possess some additional non-trivial properties, being a subset of the initial (not exponentiated)
series of diagrams. Such an approach can be considered as an exact resummation of the initial diagram series. In this way, the knowledge of the
rules for exponentiated diagrammatic significantly simplifies analysis of the Wilson lines correlators within the perturbation theory.

The exponentiation property for Abelian gauge theories has been understood a long time ago \cite{Yennie:1961ad}. The exponentiation for a
non-Abelian gauge theory is significantly more complicated, and until the recent time, only the exponentiation property for the cusp
configuration of Wilson lines and for a Wilson loop was known \cite{Sterman:1981jc,Gatheral:1983cz,Frenkel:1984pz}. The exponentiated diagrams
for the cusp were called ``webs'' \cite{Sterman:1981jc}, and were investigated in great details. They play an important role in the description
of the strong interaction dynamics. However, nowadays more complex configurations of Wilson lines are interesting. During the last decade
several approaches to the exponentiation of a general correlator of Wilson lines were presented
\cite{Laenen:2008gt,Mitov:2010rp,Gardi:2010rn,Gardi:2013ita} (the brief review of these approaches is given in sec.\ref{sec:review}).

In this article we present and elaborate details of the exponentiation approach via the generating function, which has been partially presented
in \cite{Vladimirov:2014wga}. The main object of our interest is the correlator of any number of Wilson lines of any paths and of any group
representations
\begin{eqnarray}\label{intro:general_correlator}
\mathcal{S}_N\(\{\gamma_1,f_1\},...,\{\gamma_N,f_N\}\)=\big\langle \Phi^{f_1}_{\gamma_1}\times...\times \Phi^{f_N}_{\gamma_N}\big\rangle,
\end{eqnarray}
where $\Phi$ is a Wilson line of representation $f$ along the path $\gamma$:
\begin{eqnarray}\label{into:WL_def}
\Phi^f_\gamma=P\exp\(ig\int_0^1 d\tau~\dot \gamma^\mu(\tau)A^a_\mu(\gamma(\tau))t^{[f]}_a\),
\end{eqnarray}
where $\gamma(\tau)$ parameterizes the path, and $\dot \gamma$ is a tangent to the path at point $\tau$, $t^{[f]}$ are the generators of the
gauge group in the representation $f$. We stress that the Wilson line (\ref{into:WL_def}) is a matrix in the color space, and hence, the
correlator (\ref{intro:general_correlator}) is a multi-matrix. The disentangling of the matrix structure is the main difficulty of the
exponentiation in a non-Abelian gauge theory. In particular, inaccurate treatment of the matrix structure gives rise to the incomplete
exponentiated expression in ref.\cite{Vladimirov:2014wga}. This issue is corrected in the present article.

The method of exponentiation presented  in the article, which we call the generating function exponentiation, is novel and qualitatively differs
from the methods presented in the literature. The main feature of the generating function exponentiation is that the argument of exponent is
presented in closed form, namely, as a function of correlators of certain operators. This function has the meaning of the generating function
for web diagrams.

Technically, the main difference between the generating function exponentiation and other methods of exponentiation consists in the splitting of
the problem onto two principally different mathematical tasks: the exponentiation of a scalar operator in a non-Abelian field theory, and the
exponentiation of a matrix object. Separately these tasks are effortless. However, together they produce many various diagrams and factors, the
straightforward exponentiation of which is a cumbersome work. Therefore, despite many parallels with exponentiation methods presented in
literature, we found the final result unique. The detailed explicit comparison of the exponentiation approaches is given in
sec.\ref{sec:example_of_app}.

The article is composed as follows. In sec.\ref{sec:review} we make a brief review of exponentiation methods. The main aim of this section is to
introduce a minimal terminology. In sec.\ref{sec:generation_function_approach} we present elementary introduction to the generating function
approach to the exponentiation, in particular, we demonstrate the exponentiation in the Abelian gauge theory and the exponentiation of real
exchanges. Section \ref{sec:exponentiation_in_nonA} is the main section of the article. In this section we present the generating function
approach to the non-Abelian exponentiation and derive all the main formulae. Finally, sec.\ref{sec:example_of_app} presents a set of explicit
applications of the generating function exponentiation. In particular, in sec.\ref{sec:cusp} on the example of the two-loop cusp, we perform
diagram-by-diagram comparison of the presented approach with existing approaches, while in sec.\ref{sec:MGEW} we demonstrate the effectiveness
of the method by evaluating some configurations of light-like Wilson lines in the multi-gluon-exchange-webs approximation.

\section{Brief review of exponentiation methods}
\label{sec:review}

In the paper we present an exponentiation method for a product of Wilson lines and compare it with other exponentiation approaches presented in
literature. The methods (the one presented here and those taken from literature) have similar and distinct points. In order to clarify the
presentation let us shortly describe the exponentiation methods and introduce necessary terminology.

The most straightforward approach to the exponentiation is to consider the logarithm of the perturbative series. Performing the perturbative
expansion of the logarithm and combining diagrams together one obtains the exponentiated perturbative series. Within the composition of diagrams
many terms cancel and the remainder is called web diagrams, or webs \cite{Sterman:1981jc}. We name such an approach to the exponentiation as the
\textit{diagrammatic exponentiation}.

For the cusp and for a Wilson loop the diagrammatic approach gives the following result for the exponentiated series of diagrams. The color
factor of the diagrams (of the non-exponentiated series) should be replaced by the modified color factors \cite{Gatheral:1983cz,Frenkel:1984pz},
\begin{eqnarray}\label{intro:diag_exp}
\tr\langle \Phi\rangle=\sum_{d}C(d)\mathcal{F}(d)=\exp\(\sum_{d}\tilde C(d)\mathcal{F}(d)\),
\end{eqnarray}
where $C(d)$ and $\mathcal{F}(d)$ are the color factor and the kinematical part of a diagram $d$, respectively. The modified color factor
$\tilde C$ is obtained by the recursive procedure
\begin{eqnarray}\label{intr:mod_C_def}
\tilde C(d)=C(d)-\sum_{d'}\prod_{w\in d'}\tilde C(w),
\end{eqnarray}
where the sum runs over all decompositions of $d$ into two-Wilson-lines-irreducible subgraphs $w$ (webs). In the consequence of the expression
(\ref{intr:mod_C_def}) only the two-Wilson-lines-irreducible diagrams have non-zero modified color factors. These diagrams form the set of webs
for the cusp, or for the Wilson loop.

For a general configuration of Wilson lines the diagrammatic exponentiation has been considered in ref.\cite{Mitov:2010rp}. It has been shown
that for the general case no simple selection criterium exists, and all diagrammatic topologies enter the exponentiated series.

Another method of exponentiation was suggested in ref.\cite{Laenen:2008gt} and elaborated in \cite{Gardi:2010rn,Gardi:2011wa,Gardi:2013ita}. The
method is based on the replica trick and hence, we name it as the \textit{replica exponentiation}.

Briefly, the course of the replica exponentiation is the following. In the first step, the correlator of (replica-ordered) Wilson lines is
considered in the replicated theory, i.e. the theory that consists of $N_{\text{rep}}$ copies of the original theory (replicas). The obtained
diagrammatic expansion depends on $N_{\text{rep}}$. In the second step, the terms proportional to $N_{\text{rep}}$ are taken, and the rest terms
are discarded. These terms form the argument of the exponent in the original theory. The resulting expression can be presented in the following
form
\begin{eqnarray}\label{intr:web_mixing_def}
\langle \Phi\times \Phi...\times \Phi\rangle_{IJ}=\exp\(\sum_{d,d'}\mathcal{F}(d)R_{d,d'}C(d')\)_{IJ},
\end{eqnarray}
where $R$ is the web-mixing matrix \cite{Gardi:2010rn,Gardi:2011wa}, and $I$ and $J$ are the color multi-indices. The product $RC$ generalizes
the concept of the modified color factor of the color-singlet case (\ref{intr:mod_C_def}).

The replica exponentiation gives deeper theoretical understanding of the exponentiation procedure, and allows one to make certain general
conclusions about webs. For example, using the replica exponentiation it has been shown that color factors of the exponentiated expression
contain only the color connected combinations of diagrams \cite{Gardi:2013ita}. Recently, the method has been improved by the effective vertex
formalism \cite{Gardi:2013ita}, and in this form it was used for analysis of particular multi-loop webs \cite{Falcioni:2014pka}.

The disadvantage of the replica method is that it does not present any universal expression for the series of web diagrams. The web-mixing
matrix should be calculated at every order of the perturbative expression independently, what makes any general consideration difficult. In this
aspect, the replica exponentiation resembles the diagrammatic exponentiation for the product of Wilson lines \cite{Mitov:2010rp} (for explicit
comparison see the appendix of ref.\cite{Gardi:2011wa}).

In this article we present a novel exponentiation approach. The main distinctive feature of the approach is the presentation of the
exponentiated series in closed form, namely, as a function of correlators of known operators. This function can be interpreted as a generating
function for web diagrams \cite{Vladimirov:2014wga}. Thus, we name the presented method as the \textit{exponentiation via generating function}
or, for shortness, \textit{GF exponentiation}.

Although GF exponentiation is obtained independently from the diagrammatic exponentiation and the replica exponentiation, these approaches have
many common points. GF exponentiation can be viewed as a union of both approaches. As we show later, the generating function consists of two
parts. One of those can be easily traced in the replica method (in its formulation via effective operators \cite{Gardi:2013ita}), while another
is closely related to the diagrammatic approach. However, in contrast to the diagrammatic and the replica exponentiations, the GF exponentiation
has a visual and simple expression.

\section{Generating function approach to exponentiation}
\label{sec:generation_function_approach}

In this section we discuss the basics of GF exponentiation. The material of the section is somewhat trivial and generally can be collected from
textbooks on quantum field theory. Therefore, the main aim of this section is not to present a new material, but to introduce notations, and the
general way of approach.

We also demonstrate GF exponentiation in Abelian gauge theory. This presentation serves as a pedagogical example, which is later generalized on
the non-Abelian case. The presentation of this part closely follows the presentation in ref.\cite{Vladimirov:2014wga}.

In the last subsection, we briefly discuss the generalization of GF exponentiation on the correlators with real exchanges, which are very
important in practice.

\subsection{The foundation of exponentiation}
\label{sec:foundation}

In ref.\cite{Vladimirov:2014wga} it was shown that the quantum average of any operator, which has the form of an exponent of another operator,
can be presented as an exponent of the series of connected diagrams. In fact, this statement is a reformulation of the famous relation between
the partition function and the series of connected diagrams (\ref{exp_gen:W=lnZ}). The relation holds in any quantum field theory, therefore, in
this section we use maximally abstract notation, denoting all fields of a theory as $A$, and do not specify their quantum numbers.

The partition function of a quantum theory reads
\begin{eqnarray}
Z[J]=\int DA~e^{S[A]+\int dx ~J(x)O(x)},
\end{eqnarray}
where $O(x)$ is a composite operator, $J(x)$ is the source, and $S[A]$ is the action of theory. The diagrammatic expansion of $Z[J]$ consists of
all possible Feynman diagrams connected and disconnected, with or without insertions of operator $O$. The logarithm of the partition function
\begin{eqnarray}\label{exp_gen:W=lnZ} W[J]=\ln\frac{Z[J]}{Z[0]},
\end{eqnarray}
has the meaning of the generating function for the Green functions of operators $O$. Diagrammatically, it is given by only connected diagrams
with insertions of operator $O$. In the case of many sources, $Z[J_1,J_2,...]$, the corresponding generating function $W[J_1,J_2,...]$ is given
by all connected diagrams with all possible insertions of operators $O_1$, $O_2$, etc.

Let us consider an operator of the form
\begin{eqnarray}\label{exp_gen:O=e^Y}
O[A]=\exp\(\int dx~M(x)\mathcal{Y}[A]\),
\end{eqnarray}
where $M(x)$ is some classical field  and $\mathcal{Y}[A]$ is a composite operator of fields $A$. According to the definition of the quantum
average one can consider the vacuum matrix element of the operator (\ref{exp_gen:O=e^Y}) as the partition function evaluated on the ``classical
sources'' $M(x)$,
\begin{eqnarray}\label{exp_gen:<O>=Z}
\big\langle T~O[A]\big\rangle=\frac{1}{Z[0]}\int DA~e^{S[A]}\,e^{\int dx~M(x)\mathcal{Y}(x)}=\frac{Z[M]}{Z[0]}.
\end{eqnarray}
The T-ordering on the left-hand-side of eqn.(\ref{exp_gen:<O>=Z}) allows us to apply the functional integration quantization. For brevity, in
the following we do not explicitly denote T-ordering, assuming it for every quantum average. Exception is made only in sec.\ref{sec:real_gluon},
where we discuss the quantum averages of operators without T-ordering.

The matrix element (\ref{exp_gen:<O>=Z}) can be presented as an exponent of the generating function (\ref{exp_gen:W=lnZ}). By definition, the
generating function can be presented as a series of correlators
\begin{eqnarray}\label{exp_gen:<O>=exp(W)}
\big\langle O[A]\big\rangle&=&e^{W[M]} \\\nn&=&\exp\(\int dx M(x)\langle \mathcal{Y}(x)\rangle +\frac{1}{2}\int dx_{1,2} M(x_1)M(x_2)\langle
\mathcal{Y}(x_1)\mathcal{Y}(x_2)\rangle\right.
\\\nn &&\qquad\qquad\qquad\left.
+\frac{1}{3!} \int dx_{1,2,3} M(x_1)M(x_2)M(x_3)\langle \mathcal{Y}(x_1)\mathcal{Y}(x_2)\mathcal{Y}(x_3)\rangle+... \),
\end{eqnarray}
where the factorial coefficients are the symmetry coefficients resulting from the permutation symmetry of the correlators. Thus, the
exponentiated series for the operator $O[A]$ is given by only connected diagrams with arbitrary number of operators $\mathcal{Y}$ convoluted
with the ``classical sources'' $M(x)$.

The fact that $W[M]$ is given solely by connected diagrams does not imply that the original series for $\langle O\rangle$ contains disconnected
diagrams. Both series contain only connected diagrams, but connected to different operator vertices. Since, the operators $\mathcal{Y}$ and $O$
can have very different properties, the series of diagrams contributing to $\langle O\rangle$ can relate to the series of diagrams contributing
to $W[M]$ in a very non-trivial way. The case of non-Abelian exponentiation, the main subject of this article, is an example of such a
non-trivial relation.

Exponentiation of the perturbative series for an operator of the type (\ref{exp_gen:O=e^Y}) is the fundamental property of the perturbative
expansion. It is founded only on the  relations between the symmetry coefficients of various Feynman diagrams. In its own turn, the later is the
consequence of the perturbative approach to functional integration, i.e. the property of the expansion of the action exponent around its
Gaussian part. It is important to mention that the structure of the propagator does not affect the symmetry coefficients and, thus, does not
influence on the exponentiation of the diagrams.

\subsection{Exponentiation of Wilson lines in QED}
\label{sec:QED}

Let us demonstrate the application of GF exponentiation of the Wilson line in an Abelian gauge theory (QED).

The starting point of the approach is to present the operator, namely the Wilson line, in the form (\ref{exp_gen:O=e^Y}). In QED the
path-ordered exponent (\ref{into:WL_def}) is equal to the usual exponent,
\begin{eqnarray}
\Phi_\gamma=\exp\(ig\int_0^1 d\tau~\dot \gamma^\mu(\tau)A_\mu(\gamma(\tau))\).
\end{eqnarray}
Comparing this expression with eqn.(\ref{exp_gen:O=e^Y}) we find that the operator $\mathcal{Y}(x)$ is just the field $A_\mu(x)$, while the
source $M(x)$ is a classical source of radiation,
\begin{eqnarray}\label{exp_gen:Abel_M}
M^\mu(x)=ig\int_0^1d\tau \,\dot\gamma(\tau)\,\delta(\gamma(\tau)-x).
\end{eqnarray}

Evaluating the expression (\ref{exp_gen:<O>=exp(W)}) with the sources (\ref{exp_gen:Abel_M}) we obtain the exponentiated expression
\begin{eqnarray}\label{exp_gen:Abel_exp1}
\big\langle \Phi_\gamma\big\rangle&=&\exp\Big(\frac{-g^2}{2!}\int_0^1 d\tau_1\int_0^1
d\tau_2\dot\gamma^{\mu_1}(\tau_1)\dot\gamma^{\mu_2}(\tau_2) \big\langle A_{\mu_1}(\gamma(\tau_1))A_{\mu_2}(\gamma(\tau_2))\big\rangle
\\\nn&&+\frac{g^4}{4!}\(\prod_{i=1}^4\int_0^1d\tau_i\dot \gamma^{\mu_i}(\tau_i)\)
\big\langle A_{\mu_1}(\gamma(\tau_1))A_{\mu_2}(\gamma(\tau_2))A_{\mu_3}(\gamma(\tau_3))A_{\mu_4}(\gamma(\tau_4))\big\rangle+...\Big),
\end{eqnarray}
where the correlators with odd number of photons are omitted in consequence of Furry's theorem. Diagrammatically the argument of the exponent is
given by connected diagrams with an arbitrary number of external photons positioned on the path of the Wilson line.

The expression (\ref{exp_gen:Abel_exp1}) is already suitable for the further consideration, however, it can be rewritten in the more traditional
form. Indeed, using the symmetry of the multi-photon correlators under the permutations of fields we combine multiple sources to a single
path-ordered source,
\begin{eqnarray}\label{exp_gen:Abel_exp2}
&&\big\langle \Phi_\gamma\big\rangle=\exp\Big(-g^2\int_0^1 d\tau_1\int_{\tau_1}^1 d\tau_2\dot\gamma^{\mu_1}(\tau_1)\dot\gamma^{\mu_2}(\tau_2)
\big\langle A_{\mu_1}(\gamma(\tau_1))A_{\mu_2}(\gamma(\tau_2))\big\rangle
\\\nn&&+g^4\int_0^1\!\int_{\tau_1}^1\!\int_{\tau_2}^1\!\int_{\tau_3}^1\!\(\prod_{i=1}^4d\tau_{i}\dot \gamma^{\mu_i}(\tau_i)\!\)
\!\!\big\langle
A_{\mu_1}\!(\gamma(\tau_1))A_{\mu_2}\!(\gamma(\tau_2))A_{\mu_3}\!(\gamma(\tau_3))A_{\mu_4}\!(\gamma(\tau_4))\big\rangle+...\Big).
\end{eqnarray}
The symmetry coefficients of (\ref{exp_gen:Abel_exp1}) are canceled by factorial multipliers resulting from the ordering procedure. The
expression (\ref{exp_gen:Abel_exp2}) has the common form of the exponent of all completely connected (connected in the absence of Wilson line)
diagrams with unity symmetry coefficients.

We stress that within GF exponentiation the contour of a Wilson line plays no role. Therefore, the exponentiation property holds for any
contours, including cusped, self-crossed and disconnected. However, for the contours with singularities it is convenient to introduce
independent sources for every individual smooth segment. This allows one to reveal  the influence of the contour singularities on the diagrams
explicitly.

\subsection{Exponentiation of real exchanges}
\label{sec:real_gluon}

Diagrams with real particles exchanges play an important role in applications of quantum field theory, especially, in the consideration of final
state interactions of hard processes. Typically, on the operator level the real exchanges appear as an insertion of the complete set of
asymptotic states. Alternatively such operators can be given by the (usual) product of T-ordered operators
\begin{eqnarray}\label{realG:oXXo=oo}
\langle T~O_1[A] \sum_X|X\rangle\langle X|T~O_2[A] \rangle=\langle \(T~O_1[A]\) \(T~O_2[A]\) \rangle.
\end{eqnarray}
On the diagrammatic level it implies that diagrams describing (\ref{realG:oXXo=oo}) contain two parts. The interactions inside these parts are
presented by the Feynman propagators $\Delta_F$, while the interactions between these two parts are presented by the positive-frequency part of
the Pauli-Jordan function, $\Delta^{(+)}$.

We remind that the standard functional integration approach assumes the consideration of only T-ordered operators. The diagrammatic expansion of
the functional integral can be expressed via the differential reduction exponent (see \cite{Vasiliev})
\begin{eqnarray}\label{realG:reducig_exp}
\langle T~O[A]\rangle=\frac{1}{Z[0]}e^{\(\int \frac{dx dx'}{2} \frac{\delta}{\delta A(x)}\Delta_F(x,x')\frac{\delta}{\delta
A(x')}\)}O[A]e^{S_{int}[A]}\Big|_{A=0},
\end{eqnarray}
where $\Delta_F(x,x')$ is the Feynman propagator, $S_{int}$ is the interaction part of the action. In the reduction exponent
(\ref{realG:reducig_exp}) the T-ordering of operator is reflected through the Feynman propagator.

In fact, the reduction exponent is the general form of any diagrammatic expansion, but not only the T-ordered as it follows from the functional
integral. In order to obtain the diagrammatic expansion for (\ref{realG:oXXo=oo}) we can adjust every operator to an independent copy of a
quantum field theory. In this case, the virtual interactions are generated independently by copies of the functional integral. The real particle
exchanges between these copies can be added to diagrammatic by the extra reduction exponent with a real propagator. Formally, we have
\begin{eqnarray}\nn
\langle \(T~O_1[A]\) \(T~O_2[A]\) \rangle &=&\frac{1}{Z^2[0]}e^{\(\int \frac{\delta}{\delta A_1}\Delta^{(+)}\frac{\delta}{\delta
A_2}\)}\(e^{\(\frac{1}{2}\int \frac{\delta}{\delta A_1}\Delta_F\frac{\delta}{\delta
A_1}\)}O_1[A_1]e^{S_{int}[A_1]}\)\\&&\qquad\qquad\times\(e^{\(\frac{1}{2}\int \frac{\delta}{\delta A_2}\Delta_F\frac{\delta}{\delta
A_2}\)}O_2[A_2]e^{S_{int}[A_2]}\)\Big|_{A_{1,2}=0},
\end{eqnarray}
where we omit the arguments of fields for brevity. In this way, the product of T-ordered operators is presented as a T-ordered product of
operators in the modified theory,
\begin{eqnarray}\label{exp_real:TT->T}
\langle \(T~O_1[A]\) \(T~O_2[A]\) \rangle&=&\langle T~O_1[A]~O_2[A] \rangle_{\text{mod}}
\\\nn&=&\frac{1}{Z_{\text{mod}}[0]}\int DA_1DA_2~ O_1[A_1]O_2[A_2]~e^{S[A_1]+S[A_2]+\int A_1\Delta^{(+)}A_2}.
\end{eqnarray}
Such a trick has been applied in ref.\cite{Belitsky:1998tc} for calculation of Drell-Yan soft factor.

Let us consider two T-ordered operators of the form (\ref{exp_gen:O=e^Y}). In the course of GF exponentiation we have the chain of equalities
\begin{eqnarray}
\big\langle (T~O_1[A])(T~O_2[A])\big\rangle=\big\langle T~e^{\int M_1\mathcal{Y}_1[A_1]+\int M_2\mathcal{Y}_2[A_2]}\big\rangle_{\text{mod}}
=e^{W_\text{mod}[M_1,M_2]}.
\end{eqnarray}
The modified argument of the exponent reads
\begin{eqnarray}\label{exp_real:Wmod}
W_\text{mod}[M_1,M_2]&=&\int dx\( M_1(x)\langle T~\mathcal{Y}_1[A_1]\rangle_{\text{mod}}+M_2(x)\langle
T~\mathcal{Y}_2[A_2]\rangle_{\text{mod}}\)
\\\nn&&+\int dx_{1,2} \Big(\frac{1}{2}M_1(x_1)M_1(x_2)\langle T~\mathcal{Y}_1[A_1]\mathcal{Y}_1[A_1]\rangle_{\text{mod}}
\\\nn&&\qquad\qquad\qquad+\frac{1}{2}M_2(x_1)M_2(x_2)\langle T~\mathcal{Y}_2[A_2]\mathcal{Y}_2[A_2]\rangle_{\text{mod}}
\\\nn&&\qquad\qquad\qquad+M_1(x_1)M_2(x_2)\langle T~\mathcal{Y}_1[A_1]\mathcal{Y}_2[A_2]\rangle_{\text{mod}}\Big)+...,
\end{eqnarray}
where the arguments of operators are omitted for brevity.

Any expression in the modified theory can be transformed back to the usual theory using the rule (\ref{exp_real:TT->T}). Therefore,
eqn.(\ref{exp_real:Wmod}) transforms to
\begin{eqnarray}\label{exp_real:W}
W[M_1,M_2]&=&\int dx\( M_1(x)\langle T~\mathcal{Y}_1\rangle+M_2(x)\langle T~\mathcal{Y}_2\rangle\)
\\\nn&&+\int dx_{1,2} \Big(\frac{1}{2}M_1(x_1)M_1(x_2)\langle T~\mathcal{Y}_1\mathcal{Y}_1\rangle
+\frac{1}{2}M_2(x_1)M_2(x_2)\langle T~\mathcal{Y}_2\mathcal{Y}_2\rangle\\\nn&& \qquad\qquad\qquad\qquad\qquad\qquad+M_1(x_1)M_2(x_2)\langle
\(T~\mathcal{Y}_1\)\(T~\mathcal{Y}_2\)\rangle\Big)+...,
\end{eqnarray}
where the arguments of operators are omitted for brevity. We conclude that the diagrams with real exchanges exponentiate in the same way as the
usual virtual diagrams.

\section{Exponentiation of Wilson lines in non-Abelian gauge theories}
\label{sec:exponentiation_in_nonA}

The exponentiation in non-Abelian gauge theories is a more involved problem than the exponentiation in Abelian ones. In comparison to QED there
are two sources of complication. The first complication comes from the involved form of operators $\mathcal{Y}$. The second complication comes
from the (color-)matrix structure of the non-Abelian Wilson line. The structure of operators $\mathcal{Y}$ and the generating function was
elaborated in ref.\cite{Vladimirov:2014wga}. However, in ref.\cite{Vladimirov:2014wga} the matrix issues of the non-Abelian exponentiation were
missed. In this section we present a detailed derivation of the non-Abelian exponentiation.

We remind that within the framework of GF exponentiation the only important information is the path-order of gauge fields, but not their
coordinates. The term Wilson line, which is used all over the article, denotes rather a Wilson curve on arbitrary path, than a straight Wilson
line, as it may be suggested from the term. Generally, the path can be arbitrary difficult, or even non-analytical. However, we suppose that
every individual Wilson line is smooth.

\subsection{Wilson line as exponent}
\label{sec:Log_of_Wilson}

The starting point of GF exponentiation is to present the Wilson line (\ref{into:WL_def}) in the form (\ref{exp_gen:O=e^Y}). For this purpose we
use the following exponential representation for the Wilson line,
\begin{eqnarray}\label{nonA_exp:W=expY}
\Phi^f_\gamma&=&\exp\Bigg\{ig\int_0^1 A_0+\sum_{s=1}^\infty(ig)^{s+1}\sum_{k=1}^s\frac{(-1)^k}{k+1}\\\nn
&&\times\sum_{\substack{j_1+..+j_k=s \\
j_i\geqslant 1}}\int_0^1\(\int_0^\tau...\int_0^{\tau_{j_1-1}}\text{ad}_{A_1}...\text{ad}_{A_{j_1}}\)
...\(\int_0^\tau...\int_0^{\tau_{j_k-1}}\text{ad}_{A_1}...\text{ad}_{A_{j_k}}\) A_0\Bigg\},
\end{eqnarray}
where $A_i=\dot \gamma^\mu(\tau_i)\,\hat A_\mu(\gamma(\tau_i))\,d\tau_i$ and $\tau_0=\tau$, with $\hat A_\mu(x)=t_a^{[f]}A^a_\mu(x)$. The
operator $\text{ad}_A$ is defined as $\text{ad}_AX=[A,X]$. The detailed derivation of this relation can be found in ref.
\cite{Methods_of_noncommutative_analysis}

The representation (\ref{nonA_exp:W=expY}) reveals several important properties. The main of them is that the operator $\mathcal{Y}$ consists
only of completely nested commutators of generators of gauge fields. It leads to the color-connectivity of webs, the property that has been the
defining property of the webs for many years starting from \cite{Gatheral:1983cz}.

The completely nested commutator structure of the operators also implies universality of the representation (\ref{nonA_exp:W=expY}) for Wilson
lines of different group representation. Indeed, the commutator of generators is proportional to the structure constant
$[t^{[f]}_a,t^{[f]}_b]=if_{abc}t_c^{[f]}$. The structure constant is independent of group representation. Therefore, the representation
dependence of the operator $\mathcal{Y}$ is concentrated in a single generator, which can be moved out of the operator. The equation
(\ref{nonA_exp:W=expY}) takes the form
\begin{eqnarray}\label{nonA_exp:W=exptV}
\Phi^f_\gamma=\exp\(t_a^{[f]}V^a_\gamma\),
\end{eqnarray}
where $V_\gamma^a$ are operators independent on the representation of the Wilson line. Here the definition of the operator $V$ is slightly
different from the corresponding definition in ref.\cite{Vladimirov:2014wga}, where the corresponding operator was defined with the integration
over $\tau_0$ removed. The representation (\ref{nonA_exp:W=exptV}) also has an advantage that all the matrix structure of the Wilson line is
concentrated in the single generator $t^{[f]}_a$.

In the case of multiple Wilson lines, it is convenient to merge all Wilson lines to a single exponent. With this purpose we demand that every
separate Wilson line acts in a separate matrix space. Then the whole composition of Wilson lines is a matrix of the reducible representation
$f_1\otimes f_2\otimes...\otimes f_N$, where $f_i$ is the representation of $i$'th Wilson line. This trick is often used to simplify the
consideration of web diagrams, see e.g.\cite{Laenen:2008gt,Mitov:2010rp}.  In order to simplify the notation we introduce the generator in this
space with appropriate numbering
\begin{eqnarray}\label{nonAbel:T^A_def}
T_A=\left\{
\begin{array}{cl}
t_A^{[f_1]}\otimes\pmb 1\otimes..\otimes \pmb 1,&A=1,...,\text{dim}_G,
\\
\pmb 1\otimes t_A^{[f_2]}\otimes..\otimes \pmb 1,&A=\text{dim}_G+1,...,2\text{dim}_G,
\\
...&
\\
\pmb 1\otimes\pmb 1\otimes..\otimes t_A^{[f_N]},&A=(N-1)\text{dim}_G+1,...,N\text{dim}_G~,
\end{array}
\right.
\end{eqnarray}
where $\text{dim}_G$ is the dimension of the gauge group or, equivalently, the number of generators. We adopt the convention that the labels in
the joined space are capitalized, while the labels in the irreducible spaces are denoted by the lowercase letters. Simultaneously, the path
dependence of the operator $V^a_\gamma$ is adjusted to the corresponding sector of label $A$,
\begin{eqnarray}\label{nonAbel:V^A_def}
V^A=\left\{
\begin{array}{cl}
V^A_{\gamma_1},&A=1,...,\text{dim}_G,
\\
V^A_{\gamma_2},&A=\text{dim}_G+1,...,2\text{dim}_G,
\\
...&
\\
V^A_{\gamma_N},&A=(N-1)\text{dim}_G+1,...,N\text{dim}_G~.
\end{array}
\right.
\end{eqnarray}
In this representation the product of several Wilson lines reads
\begin{eqnarray}\label{nonAbel:Phi^n=Phi}
\(\Phi^{f_1}_{\gamma_1}\)_{i_1j_1}\times \(\Phi^{f_2}_{\gamma_2}\)_{i_2j_2}\times...\times \(\Phi^{f_N}_{\gamma_N}\)_{i_Nj_N}=\Phi_{IJ}= \(\exp
T_AV^A\)_{IJ},
\end{eqnarray}
where $I$ and $J$ are indices of the joined matrix space.

The representation (\ref{nonAbel:Phi^n=Phi}) resembles the required form (\ref{exp_gen:O=e^Y}), where generators $T_A$ play the role of the
matrix sources for operators $V^A$. For the general discussion of GF exponentiation it is convenient to keep the infinite sum of operators
(\ref{nonA_exp:W=expY}) as a single object $V^A$. While for the perturbative analysis and practical applications it is convenient to split the
operator $V^A$ on individual terms of a fixed perturbative order,
\begin{eqnarray}\label{nonA_exp:V=sumV}
V^A=\sum_{n=1}^\infty V_n^A,
\end{eqnarray}
where $V_n\sim g^n$. The expressions for the first few operators $V_n$, as well as, Feynman rules for them are presented in the appendix
\ref{sec:app_op}.

\subsection{Matrix exponentiation}
\label{sec:matrix_exp}

According to the general discussion of sec.\ref{sec:foundation}, the connected diagrams with all possible insertions of the operators $V^A$ are
the only diagrams contributing to the exponent. However, in the case of non-Abelian gauge theory it is not entirely correct. The point is that
in the non-Abelian gauge theories the operators in diagrams are caused by the matrix sources $T^A$. This complication spoils the usual relations
between connected and disconnected diagrams, and prevent the straightforward exponentiation.

The direct way to bring the diagrammatic series into the exponentiated form is to consider its logarithm. The perturbative expansion of the
logarithm (of the perturbative series) mixes up the diagrams such that it is very difficult to find out on a general level, which parts of
diagrams cancel and which do not. Therefore, this way necessarily leads us to the analysis of individual Feynman diagrams, the procedure which
that we try to avoid. This approach was elaborated in ref.\cite{Mitov:2010rp}, and indeed it appears to be not efficient.

In order to solve the matrix complications of the non-Abelian exponentiation in the most efficient way, we split the consideration onto two
independent tasks. First, we consider the exponentiation of a scalar version of the non-Abelian Wilson line. That grants us the most promising
starting point for the exponentiated diagrammatic. Second, we generalize the scalar version of Wilson line and its exponentiated expression on
the matrix form. In this way we obtain the complete matrix exponentiated expression without lost of efficiency of GF exponentiation.

The first point of our program is to consider a scalar operator
\begin{eqnarray}\label{nonAbel:scalarPhi_def}
\phi=\exp\(M^AV_A\),
\end{eqnarray}
where $M^A$ is a scalar source. We emphasize that the term scalar in this section denotes an object, which does not carry matrix indices $(IJ)$,
although the object can be a vector, or a tensor in the color space. For example, the operator $V^a_\gamma$ in eqn.(\ref{nonA_exp:W=exptV}) is a
scalar, while the generator $t^a_{ij}$ is a matrix.

The operator $\phi$ has many common properties with the non-Abelian Wilson line $\Phi$. So, one may say that $\phi$ is a scalar image of the
non-Abelian Wilson line. According to the discussion of sec.\ref{sec:foundation} the average of $\phi$ can be presented in the form of exponent
\begin{eqnarray}\label{nonAbel:Wscalar_def}
\langle \phi \rangle=\frac{Z[M]}{Z[0]}=\exp W[M].
\end{eqnarray}
The function $W[M]$ is given by all connected diagrams with insertions of operator $V_A$. The exponentiation of the scalar image of Wilson lines
is, as simple as, the exponentiation of Wilson lines in the Abelian gauge theory, discussed in sec.\ref{sec:QED}. The only difference is that in
the non-Abelian gauge theory the operator $V$ is given by infinite series of operators (\ref{nonA_exp:V=sumV}).

In order to fulfill the second point of our program, we need the formal definition of the matrix generalization procedure for a function of
several arguments. It can be presented as action of the matrix shift operator on the scalar image,
\begin{eqnarray}\label{nonAble:matrix_gen_def}
\widetilde f(T)_{ij}=\(e^{T^A\frac{\partial}{\partial x^A}}\)_{ij}f(x)\Big|_{x=0}.
\end{eqnarray}
It is straightforward to check that right-hand-side of eqn.(\ref{nonAble:matrix_gen_def}) satisfies all standard demands on the matrix function
$\widetilde f(T)$.

The Wilson line $\Phi$ is obtained from its scalar image (\ref{nonAbel:scalarPhi_def}) with the help of operation (\ref{nonAble:matrix_gen_def})
\begin{eqnarray}
\Phi_{IJ}=\(e^{T^A\frac{\delta}{\delta M^A}}\)_{IJ}\phi\Big|_{M=0}.
\end{eqnarray}
Therefore, the average of Wilson line is given by
\begin{eqnarray}
\Big\langle\Phi_{IJ}\Big\rangle=\(e^{T^A\frac{\delta}{\delta M^A}}\)_{IJ}\frac{Z[M]}{Z[0]}\Bigg|_{M=0}=\frac{\widetilde Z_{IJ}[T]}{Z[0]},
\end{eqnarray}
where $\widetilde Z[T]$ is the matrix generalization of the partition function by means of the procedure (\ref{nonAble:matrix_gen_def}).

As one can see from eqn.(\ref{nonAble:matrix_gen_def}), the generalization of the partition function to matrix sources is rather straightforward
procedure. However, the relations between diagrams within $\widetilde Z[T]$ are not the same as within $Z[M]$. This happens due to the
symmetrization of matrix variables by the shift operator (\ref{nonAble:matrix_gen_def}). As a result, the originally disconnected diagrams are
entangled by their matrix structure (see explicit example in eqn.(\ref{nonAbel:example_SymmProd})). Therefore, the logarithm of the matrix
partition function $\widetilde Z[T]$ is not the matrix generalization of the function $W[M]$ (which we denote as $\widetilde W[T]$). In symbolic
notations this statement reads
\begin{eqnarray}\label{nonAbel:neq}
\frac{\widetilde Z_{IJ}[T]}{Z[0]}=\frac{1}{Z[0]}\(e^{T^A\frac{\delta}{\delta M^A}}\)_{IJ}e^{W[M]}\Big|_{M=0}\neq \(e^{\widetilde W[T]}\)_{IJ},
\end{eqnarray}
where
\begin{eqnarray}\label{nonAbel:MEK_def}
\widetilde W_{IJ}[T]=\(e^{T^A\frac{\delta}{\delta M^A}}\)_{IJ}W[M]\Big|_{M=0}.
\end{eqnarray}
The inequality (\ref{nonAbel:neq}) has been overlooked in ref.\cite{Vladimirov:2014wga}, therefore, the final result of
ref.\cite{Vladimirov:2014wga} presented there is incomplete.

Considering the left-hand-side of eqn.(\ref{nonAbel:neq}) we conclude that although the function $\widetilde W$ is not the complete result of
the exponentiation, but still it is the only function that can appear in the argument of the exponent. Taking into account that the leading term
of the perturbative expansion is $\widetilde W$, we can present the left-hand-side of eqn.(\ref{nonAbel:neq}) as
\begin{eqnarray}\label{nonA_exp:Z=exp(W+dW)}
\frac{\widetilde Z_{IJ}[T]}{Z[0]}=\(e^{\widetilde W[T]+\widetilde{\delta W}[T]}\)_{IJ},
\end{eqnarray}
where $\widetilde{\delta W}$ is a function of $\widetilde W$.

One can see that the function $\widetilde W$ plays an exceptional role in the GF exponentiation procedure. For this reason we call $\widetilde
W$ \textit{the kernel of matrix exponentiation} (MEK).

The correction terms $\widetilde{\delta W}$ we name \textit{the defect of matrix exponentiation} (for shortness we often call it \textit{the
defect}). Mathematically, the defect arises from the reordering of color-matrices during the formation of matrix exponent. Accordingly, the
structure of the defect resembles the famous tail of commutators in Baker-Campbell-Hausdorff (BCH) formula. The defect is an algebraic function
of $\widetilde W[T]$, in the same manner as the higher terms of BCH series are algebraic functions of the previous ones. The formal definition
of the defect reads
\begin{eqnarray}\label{nonAbel:defect_def}
\widetilde{\delta W}_{IJ}[T]&=&\[\ln~,\(e^{T^A\frac{\delta}{\delta M^A}}\)_{IJ}\]e^{W[M]}\Bigg|_{M=0} \\\nn&=& \ln\(\(e^{T^A\frac{\delta}{\delta
M^A}}\)_{IJ}e^{W[M]}\Bigg|_{M=0}\)-\(e^{T^A\frac{\delta}{\delta M^A}}\)_{IJ}\ln e^{W[M]}\Bigg|_{M=0}.
\end{eqnarray}

Thus, we have shown that the (T-)product of Wilson lines can be presented as a matrix exponent
\begin{eqnarray}\label{nonAbel:Phi=e^(W+dW)}
\big\langle\(\Phi^{f_1}_{\gamma_1}\)_{i_1j_1}...\(\Phi^{f_N}_{\gamma_N}\)_{i_Nj_N}\big\rangle=\big\langle\Phi_{IJ}\big\rangle=\(e^{\widetilde
W[T]+\widetilde{\delta W}[T]}\)_{IJ},
\end{eqnarray}
where functions $\widetilde W$ and $\widetilde {\delta W}$ are defined in (\ref{nonAbel:MEK_def}) and (\ref{nonAbel:defect_def}) respectively.
This expression is the main result of the article. In the following we discuss the properties of MEK and the defect, and demonstrate the
relation between the presented approach and other approaches.

The form of the exponent (\ref{nonAbel:Phi=e^(W+dW)}) would necessarily appear during the exponentiation of any matrix object. The most
important point of exponentiation is the selection of an efficient leading term, which is MEK in GF exponentiation. As we already told, in
ref.\cite{Mitov:2010rp} the whole perturbative series has been chosen as a leading term, and this choice leads to unnecessary complications. Our
choice of MEK is inspired by many remarkable features of $\widetilde{W}$. Above all, MEK is the generating function for webs in the case of the
scalar image of Wilson line. It implies that MEK already contains essential features of the exponentiated expression, except the matrix issues.
Another example of important features favoring MEK is that MEK contains all color connected diagrams (see details in
ref.\cite{Vladimirov:2014wga}). The color connectivity is a defining attribute of web diagrams \cite{Gardi:2013ita}. All in all, MEK is the best
candidate for the leading term and, as we demonstrate later, such a conjecture is well-founded.

The structure presented in eqn. (\ref{nonAbel:Phi=e^(W+dW)}) is can be also traced in the replica exponentiation
\cite{Laenen:2008gt,Gardi:2010rn,Gardi:2011wa,Gardi:2013ita}. Within the replica exponentiation the diagrams contributing to the exponent always
has two contributions: the part proportional to unity in the replica space, and the rest, which has more involved structure (for explicit
examples see e.g. equations (50),(61),(63) and (65) in \cite{Gardi:2013ita}). The part proportional unity, after the summation over replica
indices, is linear in $N_{\text{rep}}$, and therefore, directly contributes to the exponent. In GF exponentiation this part is given by MEK. The
rest is an arbitrary polynomial of $N_{\text{rep}}$, the linear part of which contributes to the exponent. In GF exponentiation this
contribution is given by the defect. We emphasize that in contrast to the replica exponentiation, in GF exponentiation the defect
(\ref{nonAbel:defect_def}) is given by explicit expression, that significantly simplify the consideration of webs. The detailed comparison of
the approaches, as well as, an example-calculation are given in sec.\ref{sec:example_of_app}.

\subsection{Structure of MEK}
\label{sec:MEK}

MEK is a matrix generalization of the function $W[M]$ (\ref{nonAbel:Wscalar_def}). Using the explicit expression for generating function
(\ref{exp_gen:<O>=exp(W)}) we obtain
\begin{eqnarray}\label{nonAbel:tildeW[M]=1+2+3+}
\widetilde W_{IJ}[T]=T_{IJ}^A\big\langle V^A\big\rangle+\frac{\(T^AT^B\)_{IJ}}{2!}\big\langle
V^AV^B\big\rangle+\frac{\(T^AT^BT^C\)_{IJ}}{3!}\big\langle V^A V^B V^C\big\rangle+...~,
\end{eqnarray}
where we have used that the correlators are symmetric over permutations of operators.

Important to note that in the case of real gluon exchanges the matrix generalization of generating function is more involved. The product of
T-ordered operators is not symmetric under permutations. Therefore, the expression (\ref{nonAbel:tildeW[M]=1+2+3+}) would contain terms with
explicit symmetrization of generators.

Let us pass from the joined notation for multiple Wilson lines (\ref{nonAbel:T^A_def}-\ref{nonAbel:V^A_def}) to the consideration of individual
contributions. We continue to assume that generators related to a separate Wilson line act in the separate matrix spaces. So, the generator
$t_a^{[f_k]}$ is a matrix acting in the space $(i_k,j_k)$ and, MEK is a multimatrix in spaces $(i_k,j_k)$ for $k=1$ to $N$. For brevity we
continue to denote the multi-index by capital letters, i.e. $W_{i_1..i_N,j_1..j_N}=W_{IJ}$. We have
\begin{eqnarray}\label{nonAbel:TV=tV}
T_{IJ}^AV^A=\sum_{k=1}^N \(t_a^{[f_k]}\)_{i_kj_k}V^a_{\gamma_k},
\end{eqnarray}
where the missed indices on the right-hand-side of the expression should be given by the unity matrices. Also we will omit the representation
indication $[f]$ on generators, since it is fixed by denoting the matrix indices.

In the case of multiple Wilson lines, MEK reads
\begin{eqnarray}
\widetilde W_{IJ}&=&\sum_{k=1}^Nt^a_{i_kj_k}\Big\langle V^a_{\gamma_k}\Big\rangle+\sum_{\substack{k,l=1\\k\neq
l}}^N\frac{t^a_{i_kj_k}t^b_{i_lj_l}}{2!}\Big\langle
V_{\gamma_k}^aV_{\gamma_l}^b\Big\rangle+\sum_{k=0}^N\frac{\(t^at^b\)_{i_kj_k}}{2!}\Big\langle V_{\gamma_k}^aV_{\gamma_k}^b\Big\rangle\nn
\\
&&+\sum_{\substack{k,l,m=1\\k\neq l\neq m}}^{N}\frac{ t^a_{i_kj_k}t^b_{i_lj_l}t^c_{i_mj_m} }{3!}\Big\langle V_{\gamma_k}^a V_{\gamma_l}^b
V^c_{\gamma_m}\Big\rangle+ \sum_{\substack{k,l=1\\k\neq l}}^{N}\frac{ \(t^at^b\)_{i_kj_k}t^c_{i_lj_l}}{2!}\Big\langle V_{\gamma_k}^a
V_{\gamma_k}^b V^c_{\gamma_l}\Big\rangle \\&&\nn + \sum_{k=1}^{N}\frac{ \(t^at^bt^c\)_{i_kj_k}}{3!}\Big\langle V_{\gamma_k}^a V_{\gamma_k}^b
V^c_{\gamma_k}\Big\rangle+ ...~.
\end{eqnarray}
This can be simplified by the fixation of the order of individual Wilson line within correlator,
\begin{eqnarray}\nn
\widetilde W_{IJ}&=&\sum_{k=1}^Nt^a_{i_kj_k}\Big\langle V^a_{\gamma_k}\Big\rangle+\sum_{\substack{k,l=1\\k<
l}}^Nt^a_{i_kj_k}t^b_{i_lj_l}\Big\langle V_{\gamma_k}^aV_{\gamma_l}^b\Big\rangle+\sum_{k=1}^Nt^{\{ab\}}_{i_kj_k}\Big\langle
V_{\gamma_k}^aV_{\gamma_k}^b\Big\rangle
\\\label{nonA_exp:MEK=V+VV+VVV}
&&+\sum_{\substack{k,l,m=1\\k< l< m}}^{N}t^a_{i_kj_k}t^b_{i_lj_l}t^c_{i_mj_m}\Big\langle V_{\gamma_k}^a V_{\gamma_l}^b
V^c_{\gamma_m}\Big\rangle+ \sum_{\substack{k,l=1\\k< l}}^{N}t^{\{ab\}}_{i_kj_k}t^c_{i_lj_l}\Big\langle V_{\gamma_k}^a V_{\gamma_k}^b
V^c_{\gamma_l}\Big\rangle
\\&&+\sum_{\substack{k,l=1\\k< l}}^{N}t^a_{i_kj_k}t^{\{bc\}}_{i_lj_l}\Big\langle V_{\gamma_k}^a V_{\gamma_l}^b
V^c_{\gamma_l}\Big\rangle + \sum_{k=1}^{N}t^{\{abc\}}_{i_kj_k}\Big\langle V_{\gamma_k}^a V_{\gamma_k}^b V^c_{\gamma_k}\Big\rangle+ ...~,\nn
\end{eqnarray}
where $t^{\{a..b\}}$ denotes the product of generators symmetrized over the indices
\begin{eqnarray}\label{symmetrization_def}
t^{\{a_1...a_n\}}=\frac{1}{n!}\sum_{\sigma=\text{perm}[a_1...a_n]}t^{\sigma_1}...t^{\sigma_n}.
\end{eqnarray}

According to eqn.(\ref{nonA_exp:MEK=V+VV+VVV}) one has a simple diagrammatic rule: MEK consists of all diagrams connecting arbitrary number of
operators $V$ located on Wilson lines with unity symmetry coefficients. Every operator should be convoluted with the gauge group generator of
the corresponded Wilson line. If there are several operators adjusted to the same Wilson line they should be convoluted with the symmetrized
product of generators.

\begin{figure}[t]
\centering
\includegraphics[width=0.75\textwidth]{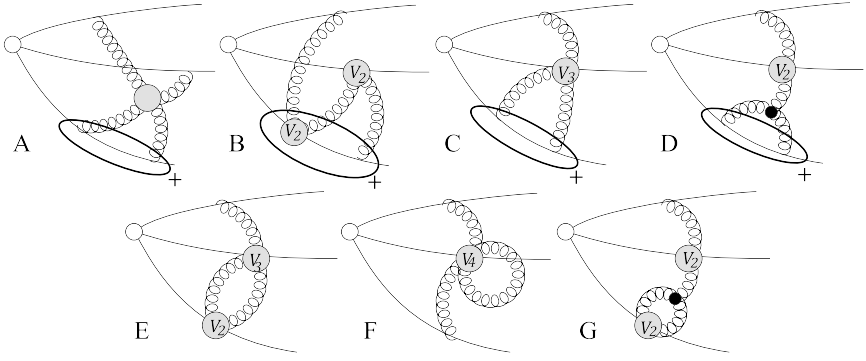}
\caption{\label{fig:MEK_example}Diagrams contributing to MEK for three Wilson lines at $\mathcal{O}(g^6)$ order that connect all three Wilson
lines (diagrams with permutations of Wilson lines should be added, as well as, web diagrams of $\mathcal{O}(g^4)$ order with a virtual loop).
Blobs with $V_n$ denote the vertices (\ref{app:V1234_op}), while the empty blob denotes all possible four-gluon tree interactions. Ovals with a
``plus'' sign denote the symmetrization of the vertices.}
\end{figure}

For the actual calculation one inserts the perturbative expansion for the operator $V$ in terms of operators (\ref{nonA_exp:V=sumV}). The
operator $V_n$ radiates exactly $n$ gluons. Nonetheless, the general diagrammatic rules are still the same: the gluons could be contracted to
other vertices, interact with each other, or be contracted to the same vertex, but all vertices $V$ must be connected together. An example of
diagrams with three Wilson lines is demonstrated in fig.\ref{fig:MEK_example}.

\subsection{Structure of the defect}
\label{sec:defect}

The formal definition of the defect is given in eqn. (\ref{nonAbel:defect_def}). However, the form (\ref{nonAbel:defect_def}) is not transparent
and is hardy applicable. In this section we reorganize eqn.(\ref{nonAbel:defect_def}) in simpler form, suitable for calculations.

In order to simplify eqn.(\ref{nonAbel:defect_def}) we expand the exponent of the generating function in a series
\begin{eqnarray}\label{nonAbel:e^W=1+W+WW}
e^{W[M]}&=&1+W_1+\frac{W_2+W_1^2}{2}+\frac{W_3+3W_2W_1+W_1^3}{3!} +...\\\nn&=& \sum_{n=0}^\infty \sum_{\substack{\{k\}\\ \sum_i i\cdot
k_i=n}}\frac{W_1^{k_1}}{(1!)^{k_1}k_1!}\frac{W^{k_2}_2}{(2!)^{k_2}k_2!}...\frac{W^{k_n}_{n}}{(n!)^{k_n}k_n!},
\end{eqnarray}
where we have introduced the short notation
\begin{eqnarray}
W_k=M^{A_1}...M^{A_k}\big\langle V^{A_1}...V^{A_k}\big\rangle.
\end{eqnarray}
The action of the matrix shift exponent (\ref{nonAble:matrix_gen_def}) replaces all sources $M^A$ by the generators $T^A$ in the completely
symmetric way. This operation mixes the matrices, such that the result is not the matrix product of $\widetilde W_n$'s. For example, the first
term with mixture of matrices in eqn.(\ref{nonAbel:e^W=1+W+WW}) is
\begin{eqnarray}\label{nonAbel:example_SymmProd}
&&\(e^{T^A\frac{\delta}{\delta M^A}}\)_{IJ}W_1W_2\Big|_{M=0}
\\\nn&&\qquad\qquad\qquad=\frac{1}{3}\(T^AT^BT^C+T^AT^CT^B+T^CT^AT^B\)_{IJ}\big\langle V^{A}V^{B}\big\rangle\big\langle V^{C}\big\rangle,
\end{eqnarray}
where the symmetry of correlators is taken into account. One can see that this expression is not equivalent to $(\widetilde W_1\widetilde
W_2+\widetilde W_2\widetilde W_1)/2$, what is needed for the expansion of $\exp(\widetilde W)$, but contains extra terms. The extra terms are
proportional to commutators of generators and give rise to the defect of exponentiation.

Let us introduce the special notation for symmetrized products of the form (\ref{nonAbel:example_SymmProd}). We denote
\begin{eqnarray}\nn
\{W_n\}_{IJ}&=&T_{IJ}^{\{A_1...A_n\}}\big\langle V^{A_1}...V^{A_n}\big\rangle\\\label{nonAbel:wSYM_def}
\{W_kW_{n-k}\}_{IJ}&=&T_{IJ}^{\{A_1...A_n\}}\big\langle V^{A_1}...V^{A_k}\big\rangle\big\langle V^{A_{k+1}}...V^{A_n}\big\rangle, \\\nn
\{W_{k}W_{l}W_{n-k-l}\}_{IJ}&=&T_{IJ}^{\{A_1...A_n\}}\big\langle V^{A_1}...V^{A_k}\big\rangle \big\langle V^{A_{k+1}}...V^{A_{k+l}}\big\rangle
\big\langle V^{A_{k+l+1}}...V^{A_n}\big\rangle,
\\\nn &&etc.
\end{eqnarray}
where $T^{\{a..b\}}$ denotes the symmetric product of generators, see (\ref{symmetrization_def}).

The operation (\ref{nonAbel:wSYM_def}) does not have a special name, nonetheless it often appears in course of diagrammatic resummations and
related problems. In ref.\cite{Mitov:2010rp} this operation serves as a generalized product of diagrams needed for the diagrammatic
exponentiation. Authors of ref.\cite{Mitov:2010rp} shows that the result of the symmetrized product (\ref{nonAbel:wSYM_def}) has the meaning of
the sum of all joined diagrams with all mutual ordering of gluons attached to Wilson lines.

In the notation (\ref{nonAbel:wSYM_def}), the matrix generalization of the partition function reads
\begin{eqnarray}\label{nonAbel:Zgen_expression}
\widetilde Z_{IJ}[T]= \(e^{T^A\frac{\delta}{\delta M^A}}\)_{IJ}e^{W[M]}\Big|_{M=0}= \sum_{n=0}^\infty \sum_{\substack{\{k\}\\ \sum_i i\cdot
k_i=n}}\frac{\{W_1^{k_1}W^{k_2}_2...W^{k_n}_{n}\}_{IJ}}{(1!)^{k_1}(2!)^{k_2}...(n!)^{k_n}k_1!k_2!...k_n!}.
\end{eqnarray}
In the same notation MEK and the partition function (\ref{nonAbel:Zgen_expression}) are
\begin{eqnarray}\label{nonAbel:Zden=eW}
\widetilde W_{IJ}[T]=\sum_{k=1}^\infty \frac{\{W_k\}_{IJ}}{k!},\qquad \widetilde Z_{IJ}[T]=\sum_{k=0}^\infty \frac{\{\widetilde
W[T]^k\}_{IJ}}{k!},
\end{eqnarray}
where the operation $\{..\}$ for $\{\widetilde W[T]^k\}$ should be taken individually for every term of perturbative expansion.

Finally, we evaluate (\ref{nonAbel:defect_def}) and obtain the expression for the defect. For the practical applications it is convenient to
present the defect as the series
\begin{eqnarray}
\widetilde{\delta W}_{IJ}[T]=\sum_{k=2}^\infty \widetilde{\delta_kW}_{IJ}[T],
\end{eqnarray}
where $\widetilde{\delta_kW}\sim \widetilde W^k$. The first few terms are
\begin{eqnarray}\nn
\widetilde{\delta_2W}_{IJ}[T]&=&\frac{1}{2}\(\{\widetilde W^2\}_{IJ}-\(\widetilde W^2\)_{IJ}\),\\
\widetilde{\delta_3W}_{IJ}[T]&=&\frac{1}{6}\{\widetilde W^3\}_{IJ}-\frac{\(\{\widetilde W^2\}\widetilde W\)_{IJ} +\(\widetilde W\{\widetilde
W^2\}\)_{IJ}}{4}+\frac{1}{3}\(\widetilde W^3\)_{IJ},\label{nonA_exp:dW1234}
\\\nn \widetilde{\delta_4W}_{IJ}[T]&=&
\frac{\{\widetilde W^4\}_{IJ}}{4!}-\frac{\(\{\widetilde W^3\}\widetilde W\)_{IJ} +\(\widetilde W\{\widetilde
W^3\}\)_{IJ}}{12}-\frac{\(\{\widetilde W^2\}\{\widetilde W^2\}\)_{IJ}}{8} \\\nn&&+ \frac{\(\{\widetilde W^2\}\widetilde W^2\)_{IJ}+\(\widetilde
W^2\{\widetilde W^2\}\)_{IJ}+\(\widetilde W\{\widetilde W^2\}\widetilde W\)_{IJ}}{6}-\frac{\(\widetilde W^4\)_{IJ}}{4}.
\end{eqnarray}

The equations (\ref{nonA_exp:dW1234}) can be significantly simplified by means of recursion. With this aim we consider
eqns.(\ref{nonA_exp:dW1234}) as equations on the totally symmetrized products $\{\widetilde W^n\}$, as functions of $\widetilde{\delta W}$.
Substituting the solution again into eqns.(\ref{nonA_exp:dW1234}) we found a simple recursive relation for the defect
\begin{eqnarray}\label{nonA:defect_recursive}
\widetilde{\delta_nW}_{IJ}[T]&=& \frac{1}{n!}\{\widetilde W^n\}_{IJ}-\sum_{k=2}^n\frac{1}{k!}\sum_{\substack{i>1\\\sum i=n}}
\(\widetilde{\delta_{i_1}W}...\widetilde{\delta_{i_k}W}\)_{IJ},
\end{eqnarray}
where $\widetilde{\delta_{1}W}=\widetilde W$. In the next sections we present explicit calculation of the defect for particular diagrams.

In fact, the matrix structure of the defect $\widetilde{\delta_nW}$ is proportional to (at least) $(n-1)$ commutators of generators. It follows
from the recursive equation (\ref{nonA:defect_recursive}). Therefore, if matrices $T$ commute, all term of the defect are zero. In this way many
contributions to the defect are zero, especially in for diagrams that involve many Wilson lines.

Due to the property that the defect is proportional to the commutators of generators, the color factors appearing in the defect are color
connected. Together with the statement that MEK is color connected, see discussion in \cite{Vladimirov:2014wga}, it proves the non-Abelian
exponentiation theorem formulated in \cite{Gardi:2013ita}.

It is interesting to consider the gauge invariance of the exponentiated expression. Generally, the correlator (\ref{nonAbel:Phi=e^(W+dW)}) is
not gauge invariant, and therefore, there are no special properties for MEK or for the defect. However, if one considers a color singlet
configuration, or the open color indices located at spatial infinities, the correlator (\ref{nonAbel:Phi=e^(W+dW)}) is perturbative gauge
invariant (for the later case, only in non-singular gauges, see e.g. discussion in ref.\cite{Belitsky:2002sm}). Consequently, the exponentiated
expression for such a gauge-invariant configuration, is also gauge invariant. Indeed, at order $\alpha$ the statement is obvious. Given that,
the gauge invariance for higher orders can be proven by iterations. In its own turn, the gauge invariance of the exponentiated expression leads
to gauge invariance of the defect and MEK independently.

Let us sketch the proof of independent gauge invariance for MEK and the defect (given that the correlator under consideration is perturbative
gauge invariant). The proof is made by interactions, comparing order-by-order the (gauge invariant) exponentiated expression and $\widetilde
W[T]+\widetilde{\delta W}[T]$. Comparing the leading terms we observe that the leading term of MEK is gauge invariant. The next-to-leading term
is given by the sum of next-to-leading term of MEK and leading term of the defect. The leading term of the defect is gauge invariant, because it
is composed from the leading terms of MEK, see (\ref{nonA:defect_recursive}). Therefore, the next-to-leading term of MEK is gauge invariant.
Then one can construct iteration and prove that MEK and the defect are independently gauge invariant at any order.

\section{Application of GF exponentiation}
\label{sec:example_of_app}

In this section we compare the GF exponentiation with the diagrammatic and the replica exponentiations. The examples presented in the section
clarify the role of the defect and MEK in the exponentiation procedure.

In order to compare approaches we consider two half-infinite Wilson lines, i.e. the cusp. The cusp is a popular playground for the testing of
various methods. Recently, it has been evaluated up to the three loop order in the general kinematics in QCD \cite{Grozin:2014hna}. The
evaluation of exponentiated diagrams at one and two-loop order can be found in many articles, e.g. in refs.
\cite{Knauss:1984rx,Korchemskaya:1992je,Korchemsky:1987wg,Erdogan:2011yc} one can find all necessary details and the explicit expressions. In
the course of comparison we do not evaluate loop-integrals. Instead, we compare the unintegrated expressions for the diagrams and show the
relations between different approaches.

In order to demonstrate the effectiveness of GF exponentiation we consider the special class of diagrams called the multiple gluon exchange webs
(MGEW). These are the diagrams with neglected interaction between gluons. Such an approximation is simplest but non-trivial part of the
exponentiated expression. Particulary, the simple structure of MGEWs results to the possibility to evaluate MGEW-loop-integrals at high orders
without significant efforts, see e.g.\cite{Henn:2013wfa,Gardi:2013saa}. MGEWs were studied in details within the replica exponentiation in
refs.\cite{Falcioni:2014pka,Gardi:2013saa}. For simplicity, we consider the configuration of light-like Wilson lines, and show that in such
kinematics MEK is given solely by one-loop diagrams. Therefore, all the higher order contribution are generated by the defect, and can be
evaluated by algebraic manipulations.

\subsection{Cusp at two loops}
\label{sec:cusp}
\begin{figure}[t]
\centering
\includegraphics[width=0.9\textwidth]{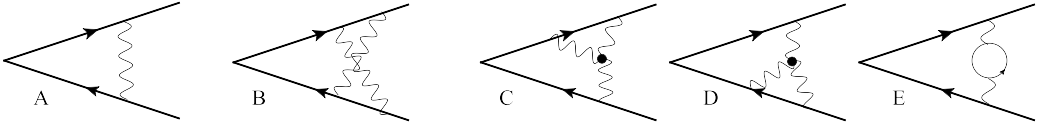}
\caption{The web diagrams contributing to the correlator of two Wilson lines. The diagram B should be taken with the modified color
factor.}\label{fig:cusp_standard}
\end{figure}
\begin{figure}[t]
\centering
\includegraphics[width=0.9\textwidth]{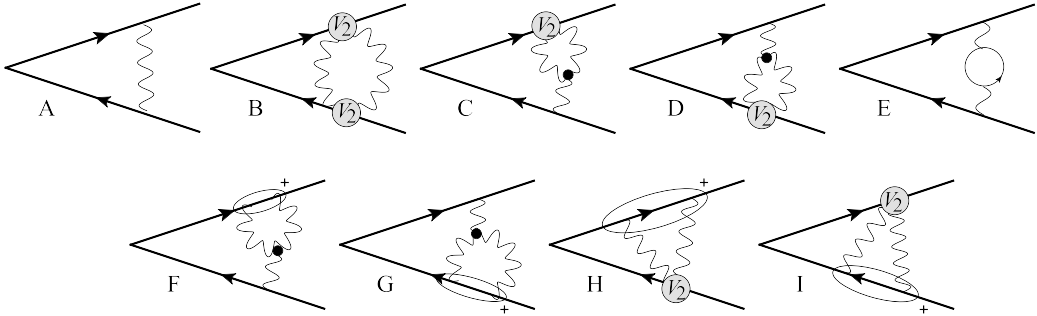}
\caption{The diagrams contributing to MEK for the cusp. The diagram B does not equal the diagram B in fig.\ref{fig:cusp_standard}. Diagrams C, D
and E, as well as, the diagram A are equal to the corresponding diagrams in fig.\ref{fig:cusp_standard}. Diagrams  F, G, H and I are zero due to
the convolution of antisymmetric three-gluon vertex with the symmetrized product of generators.}\label{fig:cusp_MEK}
\end{figure}

Let us consider a cusped Wilson line, which approaches from infinity to the origin along vector $v_1$, and then continues to infinity along
vector $v_2$. For definiteness, we consider the Wilson line in the fundamental representation of $SU(N_c)$. We denote
\begin{eqnarray}
\Big\langle \(\Phi^\dagger_{v_1}\Phi_{v_2}\)_{ij}\Big\rangle=\delta_{ij}\Gamma(v_1,v_2)=\delta_{ij}\exp\[C_F E(v_1,v_2)\],
\end{eqnarray}
where $\Phi_{v}$ is the half-infinite Wilson line pointed from the origin along the vector $v$. For simplicity, we neglect self-interaction of
Wilson lines, and use the Feynman gauge.

The webs contributing to $E$ within the diagrammatic exponentiation \cite{Gatheral:1983cz,Frenkel:1984pz} are presented on
fig.\ref{fig:cusp_standard}. The webs are to be equipped with modified color factors (\ref{intr:mod_C_def}). At this order of accuracy, the only
diagram, whose color factor differs from the original, is diagram B. Its color factor is $-C_FC_A/2$.

In order to apply GF exponentiation as it is described in sec.\ref{sec:generation_function_approach}, we consider both Wilson lines in separate
matrix spaces, see sec.\ref{sec:Log_of_Wilson}. At two-loop accuracy, only one combination of generators appears in the exponent. We denote
\begin{eqnarray}\label{cusp:Phi=exp(t1_t2_E)}
\Big\langle \(\Phi^\dagger_{v_1}\)_{i_1j_1}\(\Phi_{v_2}\)_{i_2j_2}\Big\rangle=\exp\[(t^C)^a_{i_1j_1}t^b_{i_2j_2} E^{ab}(v_1,v_2)+...\],
\end{eqnarray}
where dots stay for other compositions of color indices arise at three-loop order (see example in the next section). Here, the exponent is the
matrix exponent, where generators $t_{i_1j_1}$and $t_{i_2j_2}$ act in different matrix spaces.

The label $C$ (as if charge-conjugation) on the first generator in eqn.(\ref{cusp:Phi=exp(t1_t2_E)}) recalls that these generators should be
multiplied in the reverse order, from right to left. Such a prescription is needed to match out definition of the operator $V_\gamma$, which
utilizes the path-ordering from the origin, but not from infinity. In other words, the proper color and path ordering for conjugated Wilson line
is $\Phi^\dagger =\exp\(-t^CV\)$, where the minus sign is the result of the conjugation of operator $V$.

Contracting eqn.(\ref{cusp:Phi=exp(t1_t2_E)}) with $\delta_{j_1i_2}$ and using that function $E^{ab}$ is proportional to $\delta^{ab}$, we
obtain the relation (at this order of accuracy)
\begin{eqnarray}\label{cusp:E=E}
E^{ab}(v_1,v_2)=\delta^{ab}E(v_1,v_2).
\end{eqnarray}
In its own turn, the function $E^{ab}$ is given by the sum of MEK and the defect (\ref{nonAbel:Phi=e^(W+dW)}).

Let us remark, that one can consider the cusped Wilson line in another way. Instead of consideration of every Wilson line in separate color
space, as it is done in (\ref{cusp:Phi=exp(t1_t2_E)}), one can  consider a single Wilson line, whose path-dependent part is split,
$V^a_{\gamma}=V^a_{v_1}+V^a_{v_2}$. Within such consideration, one can keep a single generator $t^a$ in (\ref{nonA_exp:W=exptV}) by the price of
two operators $V$. These two approaches are equivalent to each other, although diagrams would distribute differently within the perturbative
expansion of MEK (\ref{nonA_exp:MEK=V+VV+VVV}).

In the following we perform the diagram-by-diagram comparison for both sides of (\ref{cusp:E=E}). The diagrams contributing to MEK are presented
in fig.\ref{fig:cusp_MEK}. In the context of eqn.(\ref{nonA_exp:MEK=V+VV+VVV}), the diagrams from A to E are produced by the perturbative
expansion of the double correlator, while the diagrams F and G are produced by the triple correlator. Contribution of the average of an operator
$V$ is zero in the absence of self-interaction.

Let us compare diagrams one-by-one. The comparison is simpler to perform in the position representation. The gluon propagator reads
\begin{eqnarray}
\Delta^{\mu\nu}_{ab}(x,y)=g^{\mu\nu}\delta_{ab}\Delta(x,y)
=\frac{\Gamma(1-\epsilon)}{4\pi^{2-\epsilon}}\frac{-g^{\mu\nu}\delta_{ab}}{(-(x-y)^2+i0)^{1-\epsilon}},
\end{eqnarray}
where $\epsilon$ is the parameter of the dimension regularization, $d=4-2\epsilon$. The expressions for the vertices $V_n$ are presented in
appendix \ref{sec:app_op}.

The diagrams A and E stay the same in both approaches. For the following consideration we need the one-loop expression for MEK,
\begin{eqnarray}\label{cusp:MEK_1loop}
\widetilde W_{\text{1-loop}}=(t^C)^a_{i_1j_1}t^b_{i_2j_2}W^{ab}_{\text{1-loop}}=-(ig)^2(t^C)^a_{i_1j_1}t^a_{i_2j_2}v_{12}\int_0^\infty dx dy
~\Delta(v_1x,v_2y),
\end{eqnarray}
where $v_{ij}=(v_i\cdot v_j)$.

The diagrams C and D are also equal in both approaches. Let us demonstrate it explicitly. The diagram C in fig.\ref{fig:cusp_standard} reads
\begin{eqnarray}
\delta_{ij}C_F E_{C}=(ig)^3\(t^at^bt^c\)_{ij}if^{abc}\mathcal{F}_{C}=\delta_{ij}C_F\frac{iC_A}{2}\mathcal{F}_{C}.
\end{eqnarray}
The kinematical part equals to
$$
\mathcal{F}_{C}=\int_0^\infty dy dx_{1,2}\int d^dz~V_{\mu\nu\rho}(z)v_1^\mu v_1^\nu v_2^\rho
\Delta(z,v_1x_1)\Delta(z,v_1x_2)\Delta(z,v_2y)\theta(x_2<x_1),
$$
where $V^{\mu\nu\rho}(z)$ is the Feynman rule for the triple-gluon vertex. The diagram C in fig.\ref{fig:cusp_MEK} reads
\begin{eqnarray}\label{cusp:MEK_C}
E^{ab}_{C}=\frac{1}{2}\frac{(i g)(ig^2)}{2}\(f^{adc}-f^{acd}\)if^{cdb}\mathcal{F}_{C}=\delta^{ab}\frac{iC_A}{2}\mathcal{F}_{C}
\end{eqnarray}
where we have used the Feynman rules for $V_2$ (\ref{app:V1234_explicit}) and the symmetry properties of the three-gluon vertex. The factor
$1/2$ in front of expression (\ref{cusp:MEK_C}) is the symmetry coefficient of the diagram. Therefore, we find  that $E_C^{ab}=\delta^{ab}E_C$.
The consideration of the diagram D is analogous.

The diagrams F, G, H and I have not analogues in the diagrammatic exponentiation. These diagrams equal zero, due to the convolution of the
symmetric combination of generators with the anti-symmetric color structure of the three-gluon vertex, or $V_2$ vertex. Explicitly, we have
\begin{eqnarray}
(t^C)^a_{i_1j_1}t^b_{i_2j_2} E^{ab}_F(v_1,v_2)=(ig)^3\(t^{\{ab\}}\)_{i_1j_1}t^{c}_{i_2j_2}if^{abc}\mathcal{F}_{F}=0,
\end{eqnarray}
where the kinematical part is
$$
\mathcal{F}_{F}=\int_0^\infty dy dx_{1,2}\int d^dz~V_{\mu\nu\rho}(z)v_1^\mu v_1^\nu v_2^\rho \Delta(z,v_1x_1)\Delta(z,v_1x_2)\Delta(z,v_2y).
$$
The consideration of diagrams G, H and I is analogous.

Finally, the diagrams B in fig.\ref{fig:cusp_standard} and in fig.\ref{fig:cusp_MEK} are not equal. This is the manifestation of the matrix
origin of Wilson lines. Within the diagrammatic exponentiation the diagram B reads
\begin{eqnarray}\label{cusp:E_B_standard}
E_B&=&-(ig)^4\frac{C_A}{2}v_{12}^2
\\\nn&&\times\int_0^\infty dx_{1,2}dy_{1,2}\Delta(v_1x_1,v_2y_1)\Delta(v_1x_2,v_2y_2)\theta(x_1<x_2)\theta(y_2<y_1).
\end{eqnarray}
While, in GF exponentiation the contribution of the diagram B to MEK reads
\begin{eqnarray}\label{cusp:E_Emek}
E^{ab}_{B,\text{MEK}}&=&\frac{-1}{2}\(\frac{ig^2}{2}\)^2f^{acd}f^{bcd}v_{12}^2 \int_0^\infty
dx_{1,2}dy_{1,2}\\&&\nn\times\Delta(x_1,y_1)\Delta(x_2,y_2)\(\theta(x_1<x_2)-\theta(x_2<x_1)\)\(\theta(y_1<y_2)-\theta(y_2<y_1)\).
\end{eqnarray}
The factor $1/2$ in front of the expression is the symmetry coefficient.

Expressions (\ref{cusp:E_B_standard}) and (\ref{cusp:E_Emek}) differ significantly. The difference is the defect of exponentiation, which comes
from the squaring of the one-loop contribution (\ref{cusp:MEK_1loop}). Evaluating the expression (\ref{nonA_exp:dW1234}) we obtain
\begin{eqnarray}\nn
\widetilde{\delta_2 W}&=&\frac{1}{2}\[(t^{\{a_1a_2\}})^C_{i_1j_1}t^{\{b_1b_2\}}_{i_2j_2}-(t^{a_2} t^{a_1})^C_{i_1j_1}(t^{b_1}
t^{b_2})_{i_2j_2}\]W_{\text{1-loop}}^{a_1b_1}W_{\text{1-loop}}^{a_2b_2}
\\&=&\frac{(ig)^4}{2}\[\frac{(t^a t^b+t^b t^a)^C_{i_1j_1}}{2}\frac{(t^a t^b+t^b t^a)_{i_2j_2}}{2}-(t^b t^a)^C_{i_1j_1}(t^a
t^b)_{i_2j_2}\]v_{12}^2\\\nn&&\qquad\qquad\qquad\qquad\qquad\qquad\qquad\times \int_0^\infty d x_{1,2}
dy_{1,2}\Delta(v_1x_1,v_2y_1)\Delta(v_1x_2,v_2y_2).
\end{eqnarray}
Simplifying the color structure we find
\begin{eqnarray}\label{cusp:E_Edef}
E^{ab}_{B,\text{defect}}&=&\frac{(ig^2)^2}{8}f^{acd}f^{bcd}v_{12}^2 \int_0^\infty dx_{1,2}dy_{1,2}\Delta(v_1x_1,v_2y_1)\Delta(v_1x_2,v_2y_2).
\end{eqnarray}
Adding the defect contribution (\ref{cusp:E_Edef}) to MEK contribution (\ref{cusp:E_Emek}) and simplifying the integrands we obtain
\begin{eqnarray}
E^{ab}_{B,\text{MEK}}+E^{ab}_{B,\text{defect}}=\delta^{ab}E_B.
\end{eqnarray}
Therefore, the sum of diagram B for MEK and the defect reproduces the diagram B within the diagrammatic exponentiation.

Thus, we observe the complete agreement between GF exponentiation and the diagrammatic exponentiation. We stress that the correspondence between
approaches is not trivial. Namely, the diagrams contributing to MEK have an unique kinematical part. Only in the combination with the defect,
the standard diagrammatic (\ref{intro:diag_exp}) is restored. It implies that GF exponentiation exponentiates not only the color part but also
the kinematic part. So, the method separates unique contributions of two-loop integrals, from the powers of one-loop contribution.

It is also instructive to compare GF exponentiation with the replica exponentiation. The diagrams appearing in replicated theory are the same as
shown in fig.\ref{fig:cusp_MEK}. The expressions to compare in the form of eqn.(\ref{cusp:Phi=exp(t1_t2_E)}) can be found in
ref.\cite{Gardi:2013ita}.

The contributions of diagrams A, C, D, E, F and G equal in both approaches. The equivalence happens because these diagrams are irreducible in
the absence of Wilson lines. Therefore, only the fields of the same replica quantum number are presented in the diagram. In other words, these
diagrams are proportional to unity in the replica space, that after evaluation of the diagrams results in the common factor $N_{\text{rep}}$
(number of replicas). Effective vertices at $N_{\text{rep}}=1$ are the same in both approaches (compare our definition of $V$ in
(\ref{nonA_exp:W=exptV}) with the definition (20) in ref.\cite{Gardi:2013ita}. Also compare explicit expressions for the vertices $V_n$
(\ref{app:V1234_op})  with (22a-22c) in ref.\cite{Gardi:2013ita}). Thus, the rest parts of diagrams are identically  equal in both approaches.

Diagrams H and I are zero in both approaches due to the convolution of symmetric and anti-symmetric color factors. In ref.\cite{Gardi:2013ita}
this fact is demonstrated in eqn.(48).

Diagram B in the replica exponentiation is presented in eqn.(52) of ref.\cite{Gardi:2013ita}. Using our notation this equation reads
\begin{eqnarray}
E^{ab}_{B,\text{rep}}&=&-\frac{(ig^2)^2}{8}f^{cda}f^{cdb}v_{12}^2 \int_0^\infty dx_{1,2}dy_{1,2} \Delta(v_1x_1,v_2y_1)\Delta(v_1x_2,v_2y_2)
\\\nn&&\times \[N_{\text{rep}}\(\theta(x_1<x_2)-\theta(x_2<x_1)\)\(\theta(y_1<y_2)-\theta(y_2<y_1)\)+N_{\text{rep}}\(N_{\text{rep}}-1\)\],
\end{eqnarray}
where we have taken into account that in ref.\cite{Gardi:2013ita} both Wilson lines are incoming. Comparing this expression with
(\ref{cusp:E_Emek}) and (\ref{cusp:E_Edef}) we observe that
\begin{eqnarray}
E^{ab}_{B,\text{rep}}=N_{\text{rep}}E^{ab}_{B,\text{MEK}}-N_{\text{rep}}(N_{\text{rep}}-1)E^{ab}_{B,\text{defect}}.
\end{eqnarray}
Therefore, the contribution of MEK and the defect are clearly distinguished within the replica exponentiation. Namely, the diagrams with only a
single copy of field (i.e. with all internal vertices proportional to $\delta_{nm}$ in replica space) are the contribution to MEK. These
diagrams are proportional to the first power of $N_{\text{rep}}$. The rest contributions are proportional to a polynomial in $N_{\text{rep}}$
and represent the contribution of the defect. The polynomial in $N_{\text{rec}}$ reflects the reordering of color matrices during
exponentiation, which is taken into account explicitly by definition (\ref{nonAbel:defect_def}) in the case of GF exponentiation.

\subsection{Multiple gluon exchange webs for light-like Wilson lines}
\label{sec:MGEW}

The webs, as well as, diagrams contributing to MEK are naturally split on subclasses by the number of connected parts of the internal graph. The
division by this principle is not gauge invariant, but it is very effective from the practical point of view.

The limiting case of such a division is the completely connected diagrams, i.e. diagrams that are connected in the absence of both Wilson lines
and vertices $V_n$ (e.g. these are diagrams A,C,D,E,F,G in fig.\ref{fig:cusp_MEK}, and diagram A in fig.\ref{fig:MEK_example}). These diagrams
are not influenced by the matrix structure of the non-Abelian gauge theory, and do not mix with the defect. For this subclass, the diagrams of
MEK are in one-to-one correspondence with the diagrams of the diagrammatic exponentiation. The limit of completely connected graphs resembles
the Abelian exponentiation procedure.

The opposite limiting case of the division is the diagrams without any internal interaction of gluons (e.g. these are diagrams B,C,E, and F in
fig.\ref{fig:MEK_example}). These diagrams have been studied in details in the recent publications \cite{Gardi:2013saa,Falcioni:2014pka}, and
are called multiple gluon exchange webs (MGEW). Within GF exponentiation MGEWs are given by the sum of MEK and the defect contributions.
Moreover, the contribution of the defect is the strongest in comparison to the more connected diagrams. In some sense, MGEWs contain the
smallest amount of a new (in comparison to the previous order of perturbative calculation) information.

MGEWs are the best demonstration of the effectiveness of GF exponentiation. In the case of light-like Wilson lines MEK can be evaluated
\textit{exactly}, and directly contributes only at one-loop order. Therefore, for light-like Wilson lines the higher-loop orders of MGEWs are
given entirely by the defect of exponentiation.

\begin{figure}[t]
\centering
\includegraphics[width=0.95\textwidth]{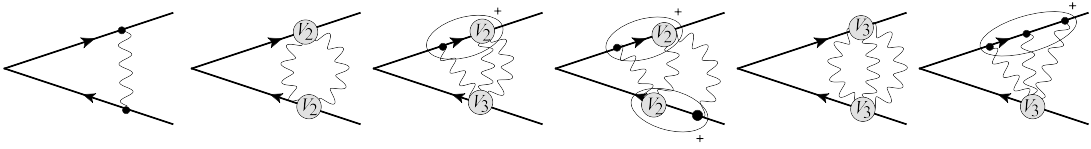}
\caption{MGEWs contributing to the cusp at three-loop level order. The symmetric diagrams are not shown. For light-like Wilson lines only the
first diagram is not zero.}\label{fig:cusp_MGEW}
\end{figure}

To start with, we consider the cusp configuration (\ref{cusp:Phi=exp(t1_t2_E)}) at $v_1^2=v_2^2=0$. MGEWs contributing to the cusp up to
three-loop level order are shown in fig.\ref{fig:cusp_MGEW}. The one-loop contribution to MEK is given by (\ref{cusp:MEK_1loop}). For the
light-like kinematic we have
\begin{eqnarray}
E_{A}^{ab}=-\delta^{ab}g^2v_{12}^{\epsilon}\frac{\Gamma(1-\epsilon)}{4\pi^{2-\epsilon}}\mu^{2\epsilon}\int_0^\infty \frac{dx dy }{(2x
y+i0)^{1-\epsilon}}e^{-\delta(x+y)},
\end{eqnarray}
where exponent performs the infrared regularization. For future convenience we rewrite the one-loop expression in the form
\begin{eqnarray}\label{MGEW:cusp_1loop}
E_{A}^{ab}=-\delta^{ab}\alpha_sv_{12}^{\epsilon}K_{1},~~~~~K_1=\frac{\Gamma^2(\epsilon)\Gamma(1-\epsilon)}{(2\pi)^{1-\epsilon}}\(\frac{\mu^2}{\delta^2}\)^\epsilon.
\end{eqnarray}

The two-loop contribution to MEK is given in eqn.(\ref{cusp:E_Emek}). In the light-like kinematics it reads
\begin{eqnarray}
E^{ab}_{B,\text{MEK}}&=&\frac{-1}{2}\(\frac{ig^2}{2}\)^2f^{acd}f^{bcd}v_{12}^{2\epsilon}
\frac{\Gamma^2(1-\epsilon)}{16\pi^{4-2\epsilon}}\int_0^\infty
dx_{1,2}dy_{1,2}\\&&\nn\times\frac{\(\theta(x_1<x_2)-\theta(x_2<x_1)\)\(\theta(y_1<y_2)-\theta(y_2<y_1)\)}{(2x_1y_1+i0)^{1-\epsilon}(2x_2y_2+i0)^{1-\epsilon}}
e^{-\delta(x_1+x_2+y_1+y_2)}.
\end{eqnarray}
The $i0$ prescription does not influence the integral, because the ultraviolet singularity at $xy\to0$ is regulated by the dimension
regularization. We conclude that the propagator attached to light-like Wilson lines can be split on individual components, $\Delta\sim
x^{\epsilon-1}y^{\epsilon-1}$. Therefore, the loop integral is zero due to the antisymmetry of the integrand under the permutation of $x_1$ and
$x_2$, or $y_1$ and $y_2$.

Thus, in the light-like kinematics of Wilson line at two-loop MGEW contribution is given solely by the defect contribution. Evaluating integral
(\ref{cusp:E_Edef}) we obtain
\begin{eqnarray}
E^{ab}_{\text{2-loop,MGEW}}=-\alpha_s^2\frac{C_A}{8}v_{12}^{2\epsilon}K_1^2.
\end{eqnarray}
This result coincides with the calculation performed in the diagrammatic exponentiation, see e.g.\cite{Erdogan:2011yc}.

In the similar way one can show that the contribution of the three- and higher-loop diagrams to MEK are zero. Indeed, all the propagators for
MGEW with light-like Wilson lines are just the product of components, $\Delta\sim x^{\epsilon-1}y^{\epsilon-1}$. Therefore, loop-integrals can
be split on two parts: one depending on the variables related to Wilson line $\Phi_{v_1}$, and another depending on the variables related to
$\Phi_{v_2}$. The interaction with Wilson lines are given by vertices $V_n$. Schematically, the contribution of some n-loop diagram reads
\begin{eqnarray}\label{MGEW:higher_loop}
\sum (\text{color factor})\!\!\int_0^\infty \!\!dx_{1,...,n}\frac{V_k(x_1,..)...V_{l}(..,x_n)}{x_1^{1-\epsilon}...\,x_n^{1-\epsilon}}
\int_0^\infty \!\!dy_{1,...,n}\frac{V_r(y_1,..)...V_{s}(..,y_n)}{y_1^{1-\epsilon}...\,y_n^{1-\epsilon}},
\end{eqnarray}
where the sets of vertices $(k,..,l)$ and $(r,..,s)$ are defined by the diagram, $k+...+l=r+...+l=n$, and the summation runs over all
independent color structures. According to the diagrammatic rules for MEK, there are necessarily several vertices $V_n$ with $n>1$. The
kinematical part of vertices $V_{n}(x_1,..,x_n)$ is antisymmetric over the permutation of arguments. Therefore, the integrals on the
right-hand-side of (\ref{MGEW:higher_loop}) are zero. The only exception is the case when all vertices are $V_1$, which is the one-loop
contribution (\ref{MGEW:cusp_1loop}).

We conclude that in the MGEW approximation, MEK is given solely by the one-loop diagram, and reads
\begin{eqnarray}\label{MGEW:cusp_MEK}
\widetilde{W}_{IJ}(v_1,v_2)\Bigg|_{\text{MGEW}}=-(t^C)^{a}_{i_1j_1}t^{a}_{i_2j_2}\alpha_sv_{12}^{\epsilon}K_{1},
\end{eqnarray}
where $K_1$ is given in (\ref{MGEW:cusp_1loop}).

The fact that at MGEW approximation MEK is given by solely by one-loop diagram, leads to the surprising conclusion. Namely, MGEW approximation
is gauge invariant (in non-singular gauges) in the case of light-like kinematics, as long as, Wilson lines are color connected at the origin.
Indeed, MEK is given by a single gauge invariant diagram. As a consequence, the defect is also gauge invariant, see discussion in the end of
sec.\ref{sec:defect}. The gauge invariance of MGEW approximation holds only for light-like Wilson lines. For the Wilson lines off-light-cone the
MEK has contributions of higher orders, which are composed from different diagrams, and therefore, MGEW diagrams are not gauge invariant.

The contribution of the defect can be obtained by straightforward evaluation of expressions (\ref{nonA_exp:dW1234}) with MEK
(\ref{MGEW:cusp_MEK}). For an arbitrary gauge group the evaluation involves the elaboration of higher Casimir operators (see e.g. four-loop
example in \cite{Henn:2013wfa}). However, for any given algebra the evaluation can be performed easily. It is convenient to present the final
result for two Wilson lines in the form
\begin{eqnarray}\label{MGEW:cusp1}
&&\Big\langle \(\Phi^\dagger_{v_1}\)_{i_1j_1}\(\Phi_{v_2}\)_{i_2j_2}\Big\rangle\Big|_{\text{MGEW}} \\\nn&&\qquad=\exp\[\sum_{n=1}^\infty
\alpha_s^n v_{12}^{n\epsilon}K_1^n \((t^C)^a_{i_1j_1}t^a_{i_2j_2}E_{n,tt}+\frac{\delta_{i_1j_1}\delta_{i_2j_2}}{N_c}E_{n,\delta\delta}\)\],
\end{eqnarray}
where $E_n$'s are numeric coefficients. Here, the exponent is the matrix exponent, where matrices with indices $(i_1j_1)$ and $(i_2j_2)$ belongs
to different matrix spaces. Often one considers the cusp configuration with contracted color indices. Then it is convenient to use the following
parametrization
\begin{eqnarray}\label{MGEW:cusp2}
\Big\langle \(\Phi^\dagger_{v_1}\Phi_{v_2}\)_{ij}\Big\rangle\Big|_{\text{MGEW}}=\delta_{ij}\exp\[C_F\sum_{n=1}^\infty \alpha_s^n
v_{12}^{n\epsilon}K_1^n E_{n,\text{cusp}}\].
\end{eqnarray}
The first few coefficients $E$ for the case of $SU(N_c)$ gauge group are presented in table \ref{table:cusp}. For the calculation we have used
the \textit{ColorMath} package \cite{Sjodahl:2012nk}.

\begin{table}[t]
\begin{center}
\begin{tabular}{|c||l|l|l|}
\hline \rule{0ex}{2.5ex} $n$ & $E_{n,tt}$ & $E_{n,\delta\delta}$ & $E_{n,\text{cusp}}$
\\[1mm]\hline
\rule{0ex}{2.5ex}$1$ & $-1$ & $0$ & $-1$
\\[1mm]\hline
\rule{0ex}{2.5ex}$2$ & $-\frac{N_c}{8}$ & $0$ & $-\frac{N_c}{8}$
\\[1mm]\hline
\rule{0ex}{2.7ex}$3$ & $\frac{3-2N_c^2}{72}$& $\frac{1-N_c^2}{48}$ & $-\frac{N_c^2}{36}$
\\[1mm]\hline
\rule{0ex}{2.5ex}$4$ & $\frac{N_c(40-33 N_c^2)}{4608}$& $\frac{5N_c(1-N_c^2)}{768}$ & $\frac{N_c(10-33 N_c^2)}{4608}$
\\[1mm]\hline
\rule{0ex}{2.9ex}$5$ & $\frac{-35+93 N_c^2-57 N_c^4}{28800}$& $\frac{(1-N_c^2)(23 N^2-22)}{23040}$ & $\frac{N_c^2(71-114 N_c^2)}{57600}$
\\[2mm]
\hline
\end{tabular}
\end{center}
\caption{The coefficients for expansion of the light-like cusp amplitude in MGEW approximation (\ref{MGEW:cusp1}-\ref{MGEW:cusp2}) for SU($N_c$)
gauge theory.} \label{table:cusp}
\end{table}

\begin{figure}[t]
\centering
\includegraphics[width=0.55\textwidth]{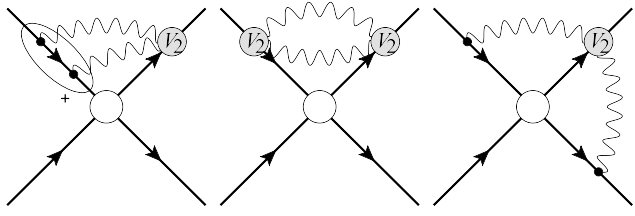}
\caption{MGEW contributing to MEK for the correlator of four Wilson lines. The diagrams with permuted Wilson lines should be added. For
light-like Wilson lines these diagrams are zero.}\label{fig:4WL_MGEW}
\end{figure}

In the case of many Wilson lines the general consideration remains the same. Since the diagrams contributing to MEK are connected in the absence
of Wilson lines they necessarily contain several vertices $V_n$ with $n>1$, see the example of two-loop MGEWs with four Wilson lines in
fig.\ref{fig:4WL_MGEW}, and three-loop MGEWs with three Wilson lines in fig.\ref{fig:MEK_example} (the diagrams B,C,E and F). The only non-zero
contribution to MEK is given by the one-loop diagrams. Therefore, MEK is given by
\begin{eqnarray}\label{MGEW:4WL_MEK}
\widetilde{W}_{IJ}(v_1,...,v_n)\Bigg|_{\text{MGEW}}=\sum_{\substack{k,l=1\\k<l}}^nt^{a}_{i_kj_k}t^{a}_{i_lj_l}\alpha_sv_{kl}^\epsilon K_{1},
\end{eqnarray}
where $K_1$ is given in (\ref{MGEW:cusp_1loop}), and missed indices should be saturated by Kronecker deltas. The expression (\ref{MGEW:4WL_MEK})
is given for the Wilson lines pointing from the origin. In the case of Wilson line incoming to the origin the corresponded generator should be
replaced by $-t^C$. Thus, the only non-trivial contribution to the correlator of light-like Wilson lines in MGEW approximation comes from the
defect and can be evaluated algebraically.

\begin{table}[t]
\begin{center}
\begin{tabular}{|c||p{14cm}|}
\hline \rule{0ex}{2.5ex} $n$ & $A_n,~B_n$
\\[2mm]\hline
\rule{0ex}{2.5ex}$1$ & $A_1=2C_Fu_1+\frac{s_1-t_1}{N_c}\qquad\qquad\qquad\qquad\qquad\qquad B_1=t_1-s_1$
\\[2mm]\hline
\rule{0ex}{2.5ex}$2$ & $A_2=\frac{C_A}{8}\(2C_Fu_2+\frac{s_2-t_2}{N_c}\)\qquad\qquad\qquad\qquad~~~~ B_2=\frac{C_A}{8}\(t_2-s_2\)$
\\[2mm]\hline
\rule{0ex}{2.5ex}$3$ & $A_3=\frac{C^2_A}{36}\Big[2C_Fu_3+\frac{3}{4N_c}\Big(\frac{7s_3}{3}+\frac{2t_3}{3}+\frac{s_2(3u_1-5t_1)}{2}
-2t_2s_1+u_2(s_1-t_1)\qquad\qquad\qquad\qquad$
{$\qquad\qquad\qquad\qquad\qquad\qquad\qquad\qquad+s_1^2(t_1-u_1)+\frac{t_1u_1^2-t_1^2u_1+s_1(t_1^2-u_1^2)}{2}+\frac{3\theta_1}{2}\Big)\Big]\qquad$}
 $B_3=\frac{C_A^2}{144}\Big[C_F\Big(8t_3-2s_3-2s_2(t_1+u_1)+6t_2(u_1-s_1)+6u_2(t_1-s_1)\qquad\qquad\qquad\qquad$
 $\qquad\qquad\qquad\qquad\qquad\qquad\qquad\qquad+3s_1(t_1^2+u_1^2)-3t_1u_1(t_1+u_1)+3\theta_1\Big)\qquad\qquad$
 $\qquad+\frac{3}{N_c}\Big(-\frac{7s_3}{3}-\frac{2t_3}{3}+2s_1t_2+\frac{s_2}{2}(5t_1-3u_1)+u_2(t_1-s_1)\qquad\qquad\qquad\qquad$
 $\qquad\qquad\qquad\qquad\qquad\qquad\qquad\qquad+(t_1-u_1)\(\frac{u_1t_1}{2}-s_1^2\)+\frac{s_1^2}{2}\(u_1^2-t^2_1\)-\frac{3\theta_1}{2}\Big)\Big]$
\\[2mm]
\hline
\end{tabular}
\end{center}
\caption{The coefficients for expansion of correlator of four light-like Wilson lines in MGEW approximation (\ref{MGEW:4WL_expression}). The
variables $s,~t,~u$ and $\theta_1$ are defined in (\ref{MGEW:stu_def}).} \label{table:4WL}
\end{table}

As an illustrative example, we present the correlator of four light-like Wilson lines in MGEW approximation in the exponentiated form. The
complete result (with eight open indices) is rather involved. Therefore, we took a particular convolution of Wilson lines, which corresponds to
the scattering in $t$-channel. In MGEW approximation such a correlator is convenient to present in the following form
\begin{eqnarray}\label{MGEW:4WL_expression}
\big\langle \(\Phi^\dagger_{v_1}\Phi_{v_3}\)_{i_1j_1}\(\Phi^\dagger_{v_2}\Phi_{v_4}\)_{i_2j_2}   \big\rangle\Big|_{\text{MGEW}}=
\qquad\qquad\qquad\qquad\qquad\qquad
\\\nn
\exp\(-2\sum_{n=1}^\infty \alpha_s^n K_1^n\[\delta_{i_1j_1}\delta_{i_2j_2}A_{n}+\delta_{i_1j_2}\delta_{i_2j_1}B_n\]\).
\end{eqnarray}
The expression for coefficients $A$ and $B$ are presented in table \ref{table:4WL} up to the three-loop order. The variables $s$, $t$ and $u$
are defined as
\begin{eqnarray}\nn
&&s_n=(v_{12})^{n\epsilon}+(v_{34})^{n\epsilon},\quad t_n=(v_{14})^{n\epsilon}+(v_{23})^{n\epsilon} \quad
u_n=(v_{13})^{n\epsilon}+(v_{24})^{n\epsilon},
\\&&\label{MGEW:stu_def}
\theta_1=(v_{12}v_{13}v_{23})^\epsilon+(v_{12}v_{14}v_{24})^\epsilon+(v_{13}v_{14}v_{34})^\epsilon+(v_{23}v_{24}v_{34})^\epsilon.
\end{eqnarray}

The calculation of higher orders in MGEW approximation does not present any difficulties, but also is not interesting. We remind that MGEW is
only an approximation, and therefore, it can be considered only as an illustration. Nonetheless, the results presented in table. \ref{table:4WL}
show some interesting features.

One can see that the two-loop expression resembles the one-loop expression. This is a known feature of the two-loop soft anomalous dimension,
see \cite{Aybat:2006wq,Aybat:2006mz}. Moreover the proportionality coefficient is the same as in MGEW approximation for the cusp, see
$E_{(1,2),\text{cusp}}$ in table \ref{table:cusp}. In fact, it is a consequence of a general statement: in two-loop approximation the soft
anomalous dimension (for light-like Wilson lines) is equal to the one-loop anomalous dimension multiplied by the two-loop cusp anomalous
dimension \cite{Aybat:2006wq,Aybat:2006mz}. This observation is known as the dipole formula for infrared singularities of scattering amplitudes
in QCD \cite{Becher:2009cu,Gardi:2009qi}. The status of the dipole formula at three-loop order and higher is currently questionable, e.g. see
ref.\cite{Dixon:2009ur}, for recent review see ref.\cite{Magnea:2014vha}.

Considering, the expression presented in table \ref{table:4WL}, we observe that at three-loop order MGEW approximation does not support the
dipole formula. Since in the considered kinematics MGEW approximation is gauge invariant, the expression given in tables \ref{table:cusp} and
\ref{table:4WL} represent some irreducible part of the complete result. In this way we conclude that the violation of the dipole formula at the
three-loop order is likely.

\section{Conclusion}

In this article we have presented an approach that allows one to evaluate the correlator of a product of Wilson lines directly in the
exponentiated form. In contrast to the existing methods of exponentiation,
e.g.\cite{Gatheral:1983cz,Frenkel:1984pz,Laenen:2008gt,Mitov:2010rp,Gardi:2010rn}, the presented method does not rely on the diagrammatical
consideration, and allows to obtain the argument of exponent in the form of correlators of known operators. To our best knowledge, it is the
first example for such a relation in a non-Abelian gauge theory.

The approach is based on the fundamental property of the perturbative expansion, namely, on the exponentiation of the connected part of Feynman
diagrams. In sec. \ref{sec:foundation} (see also \cite{Vladimirov:2014wga}) we show that this property implies the exponentiation of the
diagrammatic expansion for a wide class of operators; the operators, which can be presented in the form of the exponent (\ref{exp_gen:O=e^Y}).
Although this observation is evident and well-known, it was not regularly applied for the consideration of Wilson lines and their correlators.

In the line of derivation it was natural to split the problem of exponentiation onto two independent problems: the exponentiation within the
non-Abelian gauge theory, and the exponentiation of a matrix object (see sec. \ref{sec:matrix_exp}). As the result of such consideration the
exponentiated expression is split onto two components, those are the matrix exponentiation kernel (MEK) and the defect of exponentiation. These
two components are very different both in the form and in the content. MEK contains all the essential information about the exponentiated
expression. While the defect represents the difference between the matrix and the scalar exponentiation procedure, and it is an algebraic
function of MEK.

In this way, the knowledge of MEK results to the knowledge of the complete exponent. The expression for MEK has the form of a sum of correlators
of the operators $V$ (\ref{nonA_exp:MEK=V+VV+VVV}). The operators $V$ have very peculiar form of nested commutators of gauge fields, and play an
exceptional role in the non-Abelian exponentiation. Undoubtedly, further investigation should reveal exceptional properties of the operators
$V$, and hence, of MEK. Good illustration of non-trivial properties of MEK is the correlator of light-like Wilson lines in MGEW approximation.
In this case MEK is given exactly by one-loop diagram (see sec. \ref{sec:MGEW}).

The defect of exponentiation is an algebraic function of MEK. In sec. \ref{sec:defect} we have shown that the defect is given by the simple
recursive formula (\ref{nonA:defect_recursive}), where the most non-trivial operation is the procedure of the matrix symmetrization. Therefore,
the evaluation of the defect can be easily automated.

In sec. \ref{sec:example_of_app}, we have performed the comparison of the presented approach with existing approaches. We have shown the
complete agreement between the presented approach and the standard approach  to the cusp exponentiation at two-loop order
\cite{Gatheral:1983cz,Frenkel:1984pz}(Here, the standard approach is the approach based on the modification of color factors, which can be
applied only for singlet configurations of Wilson lines). We have also demonstrated an agreement and the relation of the presented approach to
the exponentiated diagrams obtained by the replica trick \cite{Laenen:2008gt,Gardi:2010rn}. Remarkably, the splitting of the exponent onto MEK
and the defect, can be also traced within the replica approach. Thus, we conclude that the presented approach is in complete agreement with
other methods of the exponentiation.

Using MGEW approximation we have analyzed the light-like configuration of Wilson lines. In this approximation and kinematics the web diagrams
can be evaluated with a minimal effort, they are given entirely by one-loop MEK and the corresponded defect. Remarkably, that MGEW part of the
complete expression is gauge invariant that allows us to make a judgement about the form of the whole expression. Using this approximation we
observe the violation of the dipole formula for infrared singularities of parton scattering amplitude at the three-loop order. However, we can
not strictly conclude the violation of the dipole formula due to the possible cancelation of violating terms with similar terms from the
diagrams do not contributing MGEW approximation.

The presented procedure of exponentiation allows further generalizations. So, for example, the correlator of ordinary product of Wilson lines
(i.e. the diagrammatic series with exchanges by real particles) can be exponentiated by the same procedure, as it is demonstrated in sec.
\ref{sec:real_gluon}. Therefore, the presented exponentiation procedure can be applied to a wide range of physical tasks, e.g. calculation of
the soft anomalous dimension, the threshold resummation, the evaluation of the soft factor for transverse momentum dependent factorization,
description of diffractive processes.

\acknowledgments I thank A.M.~Moiseeva and R.Pasechnik for numerous stimulating discussions. This work is supported in part by the European
Community-Research Infrastructure Integrating Activity "Study of Strongly Interacting Matter" (HadronPhysics3, Grant Agreement No. 28 3286) and
the Swedish Research Council grants 621-2011-5080 and 621-2013-4287.

\appendix

\section{Operator $V_{n}$ at low orders}
\label{sec:app_op}

In this appendix we present the explicit expression for the operators $V_n$ up to the fourth order. These expressions are used for the
calculation in the sec.\ref{sec:example_of_app}.

The operator $V^a_\gamma$ is defined in (\ref{nonA_exp:W=expY},\ref{nonA_exp:W=exptV}), while the operator $V_{n,\gamma}^a$ is defined in
(\ref{nonA_exp:V=sumV}). Elaborating the matrix structure we obtain the expressions for the first four operators
\begin{eqnarray}
V_{1}^{a}&=&ig\int_0^1 d\tau A^{a}_0,\nn\\
V_2^{a}&=&-(ig)^2\int_0^1d \tau \int_0^\tau d \tau_1 \;\tr\(t^a\[\hat A_1,\hat A_0\]\),\nn\\
V_3^{a}&=&(ig)^3\int_0^1 d \tau \int_0^\tau d\tau_1\(\frac{2}{3}\int_0^\tau-\int_0^{\tau_1}\)d\tau_2~\tr\(t^a\[\hat A_1\[\hat A_2,\hat A_0\]\]\)
\label{app:V1234_op}
\\\nn
&=& \frac{(ig)^3}{3}\int_0^1 d\tau \int_0^\tau d\tau_1\int_0^{\tau_1}d\tau_2~\tr\left\{t^a\(\[\[\hat A_0,\hat A_1\],\hat A_2\]-\[\[\hat A_1,\hat
A_2\],\hat A_0\]\)\right\},
\end{eqnarray}
\begin{eqnarray}
\nn V_4^{a}&=&-(ig)^4\int_0^1d\tau\int_0^\tau d\tau_1\(\int_0^{\tau_1}\int_0^{\tau_2}-\frac{2}{3} \int_0^{\tau_1}\int_0^\tau-\frac{2}{3}
\int_0^\tau\int_0^{\tau_2}+\frac{1}{2} \int_0^{\tau}\int_0^\tau\)d\tau_2d\tau_3~
\\\nn&&\qquad\qquad\qquad\qquad\qquad\qquad\qquad\qquad\qquad\qquad\times\tr\(t^a\[\hat A_1\[\hat A_2\[\hat A_3,\hat A_0\]\]\]\)
\\\nn&=&-\frac{(ig)^4}{6}\int_0^1d \tau \int_0^\tau d\tau_1\int_0^{\tau_1}d\tau_2\int_0^{\tau_2}d\tau_3\times
\\\nn&&
\tr\left\{t^a\(\[\[\[\hat A_1,\hat A_2\]\hat A_3\],\hat A_0\]-\[\[\[\hat A_0,\hat A_1\]\hat A_2\],\hat A_3\]+\[\[\[\hat A_0,\hat A_3\]\hat
A_2\],\hat A_1\]\right.\right.\\\nn&& \qquad\qquad\qquad\qquad\qquad\qquad\qquad\qquad\qquad\qquad \left.\left.-\[\[\[\hat A_2,\hat A_3\]\hat
A_1\],\hat A_0\]\)\right\},
\end{eqnarray}
where $\hat A_i=\dot\gamma^\mu(\tau_i)t_aA_\mu^a(\gamma(\tau_i))$ and $\hat A_0=\dot\gamma^\mu(\tau)t_aA_\mu^a(\gamma(\tau))$. Here, as
everywhere in the article, $\gamma(\tau)$ is a parametrization of a Wilson line curve, $\dot \gamma(\tau)$ is a tangent to the curve, and $t_a$
is the generator of a gauge group. The expressions for $V_{3,4}$ after the first equality symbol have the form resulting directly from eqns.
(\ref{nonA_exp:W=expY},\ref{nonA_exp:W=exptV}), while the expressions after the second equality symbol are obtained by rearranging the integrals
and renaming the integration variables.

The expression (\ref{app:V1234_op}) is valid for the Wilson line in any representation. The representation dependence is concentrated in the
generator prefactor of operator $V$ in eqn.(\ref{nonA_exp:W=exptV}). It can be archived by the selection of the proper normalization condition
for the generators and structure constants. The normalization condition used in (\ref{app:V1234_op}) is
$$\[t^a,t^b\]=if^{abc}t^c,\qquad\tr(t^at^b)=\frac{\delta^{ab}}{2},$$
for generators of any representation.

For practical application one usually consider the straight segments of Wilson lines. Let us consider a straight Wilson line from the point
$y^\mu$ to the point $z^\mu$. In this case the contour parametrization reads
\begin{eqnarray}\label{app:gamma_segment}
\gamma(\tau)=\tau z^\mu+(1-\tau)y^\mu,
\end{eqnarray}
and the tangent to the contour is the constant vector
$$
\dot \gamma^\mu(\tau)=z^\mu-y^\mu=v^\mu.
$$

The Feynman rules for the operators $V_n$ in the position space are defined as
\begin{eqnarray}
V^{~~\,\mu_1...\mu_n}_{a,a_1...a_n}(x_1,...,x_n)=\frac{\delta}{\delta A_{\mu_1}^{a_1}(x_1)}...\frac{\delta}{\delta
A_{\mu_n}^{a_n}(x_n)}V_\gamma^a\Big|_{A=0}.
\end{eqnarray}
Therefore, for a straight Wilson segment (\ref{app:gamma_segment}) the Feynman rules corresponding to operators $V_{1,2,3,4}$ read
\begin{eqnarray}\nn
V_{a,a_1}^{~\mu_1}(x_1)&=&ig\,\delta_{aa_1}v^{\mu_1}\theta_1,
\\\nn
V_{a,a_1a_2}^{~\mu_1\mu_2}(x_1,x_2)&=&\frac{ig^2}{2}f_{aa_1a_2}v^{\mu_1}v^{\mu_2}\(\theta_{12}-\theta_{21}\),
\\\label{app:V1234_explicit}
V_{a,a_1a_2a_3}^{~\mu_1\mu_2\mu_3}(x_1,x_2,x_3)&=&\frac{ig^3}{12}v^{\mu_1}v^{\mu_2}v^{\mu_3}
\\&&\nn\times\Big[
f^{aa_1b}f^{ba_2a_3}\(\theta_{123}-\theta_{132}-\theta_{231}+\theta_{321}\)+\mathcal{P} _{213}+\mathcal{P}_{312} \Big],
\\\nn
V_{a,a_1a_2a_3a_4}^{~\mu_1\mu_2\mu_3\mu_4}(x_1,x_2,x_3,x_4)&=&\frac{ig^4}{12}v^{\mu_1}v^{\mu_2}v^{\mu_3}v^{\mu_4}
\Big[f_{aa_1b}f_{ba_2c}f_{ca_3a_4}
\\\nn&&
\times\(\theta_{1234}-\theta_{1243}-\theta_{2341}+\theta_{2431}-\theta_{3214}+\theta_{3421}+\theta_{4213}-\theta_{4321}\)
\\&&\nn+\mathcal{P}_{1324}\!+\!\mathcal{P}_{1432}\!+\!\mathcal{P}_{2134}\!+\!\mathcal{P}_{2431}\!+\!\mathcal{P}_{2341}\!+\!\mathcal{P}_{3142}
\!+\!\mathcal{P}_{3412}\!+\!\mathcal{P}_{3214} \\&&\nn +\mathcal{P}_{4132}+\mathcal{P}_{4312}+\mathcal{P}_{4213} \Big],
\end{eqnarray}
where all $x_i$ are located on the path of Wilson line, and $\theta_{i..k}=\theta(y<x_i<..<x_k<z)$ the Heaviside function ordering coordinates
along the Wilson line. The symbol $\mathcal{P}_{ijk}$ stands for the first term in the bracket with permuted color indices and coordinates, i.e.
$\{a_1a_2a_3\}\to\{a_ia_ja_k\}$ and $\{x_1x_2x_3\}\to\{x_ix_jx_k\}$.

In the case of half-infinite straight Wilson lines it is convenient to use the same parametrization as for the segment
(\ref{app:gamma_segment}). The only change is that the scalar parameter runs over the infinite range $\tau\in(0,\infty)$. Such a configuration
of Wilson lines can have additional infrared divergences,  resulting from the interaction with gluons at infinity. In order to regularize these
divergences one commonly uses the suppressing exponent with infinitesimal argument \cite{Gardi:2011yz,Gardi:2013saa}. Formally such a
regularization implies the change of the gauge field within the Wilson line
\begin{eqnarray}
A_\mu(v \tau +y)~\to~A_\mu(v \tau +y)e^{-i\tau \delta\sqrt{v^2-i\varepsilon}},
\end{eqnarray}
where $\delta$ and $\varepsilon$ are infinitesimal and $\delta>\varepsilon>0$. Within such a regularization the large-distance singularities are
regularized for space-, time and light-like Wilson lines. Thus, the change of the Feynman rules for a half-infinite Wilson lines is minimal.
Namely, the function $\theta_{i..k}$ should be replaced by $\theta(y<x_i<...<x_k<\infty)\exp\(-i\delta\sqrt{v^2-i\varepsilon}\sum_i x_i\)$.


\begin{thebibliography}{99}
\bibitem{Yennie:1961ad}
  D.~R.~Yennie, S.~C.~Frautschi and H.~Suura,
  Annals Phys.\  {\bf 13} (1961) 379.

\bibitem{Sterman:1981jc}
  G.~F.~Sterman,
  AIP Conf.\ Proc.\  {\bf 74} (1981) 22.

\bibitem{Gatheral:1983cz}
  J.~G.~M.~Gatheral,
  Phys.\ Lett.\ B {\bf 133} (1983) 90.

\bibitem{Frenkel:1984pz}
  J.~Frenkel and J.~C.~Taylor,
  Nucl.\ Phys.\ B {\bf 246} (1984) 231.

\bibitem{Laenen:2008gt}
  E.~Laenen, G.~Stavenga and C.~D.~White,
  JHEP {\bf 0903} (2009) 054
  [arXiv:0811.2067 [hep-ph]].

\bibitem{Gardi:2010rn}
  E.~Gardi, E.~Laenen, G.~Stavenga and C.~D.~White,
  JHEP {\bf 1011} (2010) 155
  [arXiv:1008.0098 [hep-ph]].
\bibitem{Mitov:2010rp}
  A.~Mitov, G.~Sterman and I.~Sung,
  Phys.\ Rev.\ D {\bf 82} (2010) 096010
  [arXiv:1008.0099 [hep-ph]].

\bibitem{Gardi:2013ita}
  E.~Gardi, J.~M.~Smillie and C.~D.~White,
  JHEP {\bf 1306} (2013) 088
  [arXiv:1304.7040 [hep-ph]].

\bibitem{Vladimirov:2014wga}
  A.~A.~Vladimirov,
  Phys.\ Rev.\ D {\bf 90} (2014) 066007
  [arXiv:1406.6253 [hep-th]].

\bibitem{Gardi:2011wa}
  E.~Gardi and C.~D.~White,
  JHEP {\bf 1103} (2011) 079
  [arXiv:1102.0756 [hep-ph]].

\bibitem{Falcioni:2014pka}
  G.~Falcioni, E.~Gardi, M.~Harley, L.~Magnea and C.~D.~White,
  JHEP {\bf 1410} (2014) 10
  [arXiv:1407.3477 [hep-ph]].

\bibitem{Vasiliev}
A.~N.~Vasil'ev, \textit{ The field theoretic renormalization group in critical behavior theory and stochastic dynamics.} — CRC Press, Boca
Raton, Chapman and Hall, 2004.

\bibitem{Belitsky:1998tc}
  A.~V.~Belitsky,
  Phys.\ Lett.\ B {\bf 442} (1998) 307
  [hep-ph/9808389].

\bibitem{Methods_of_noncommutative_analysis} V.~E.~Nazaikinskii, V.~E.~Shatalov, B.~Yu.~Sternin,
\textit{Methods of noncommutative analysis : theory and applications.} - Berlin, New York: Walter de Gruyter, 1996.

\bibitem{Belitsky:2002sm}
  A.~V.~Belitsky, X.~Ji and F.~Yuan,
  Nucl.\ Phys.\ B {\bf 656} (2003) 165
  [hep-ph/0208038].

\bibitem{Grozin:2014hna}
  A.~Grozin, J.~M.~Henn, G.~P.~Korchemsky and P.~Marquard,
  Phys.\ Rev.\ Lett.\  {\bf 114} (2015) 6,  062006
  [arXiv:1409.0023 [hep-ph]].

\bibitem{Knauss:1984rx}
  D.~Knauss and K.~Scharnhorst,
  Annalen Phys.\  {\bf 41} (1984) 331.

\bibitem{Korchemskaya:1992je}
  I.~A.~Korchemskaya and G.~P.~Korchemsky,
  Phys.\ Lett.\ B {\bf 287} (1992) 169.

\bibitem{Korchemsky:1987wg}
  G.~P.~Korchemsky and A.~V.~Radyushkin,
  Nucl.\ Phys.\ B {\bf 283} (1987) 342.

\bibitem{Erdogan:2011yc}
  O.~Erdogan and G.~Sterman,
  arXiv:1112.4564 [hep-th].

\bibitem{Henn:2013wfa}
  J.~M.~Henn and T.~Huber,
  JHEP {\bf 1309} (2013) 147
  [arXiv:1304.6418 [hep-th]].
\bibitem{Gardi:2013saa}
  E.~Gardi,
  JHEP {\bf 1404} (2014) 044
  [arXiv:1310.5268 [hep-ph]].

\bibitem{Sjodahl:2012nk}
  M.~Sj\"odahl,
  Eur.\ Phys.\ J.\ C {\bf 73} (2013) 2,  2310
  [arXiv:1211.2099 [hep-ph]].

\bibitem{Aybat:2006wq}
  S.~M.~Aybat, L.~J.~Dixon and G.~F.~Sterman,
  Phys.\ Rev.\ Lett.\  {\bf 97} (2006) 072001
  [hep-ph/0606254].

\bibitem{Aybat:2006mz}
  S.~M.~Aybat, L.~J.~Dixon and G.~F.~Sterman,
  Phys.\ Rev.\ D {\bf 74} (2006) 074004
  [hep-ph/0607309].

\bibitem{Becher:2009cu}
  T.~Becher and M.~Neubert,
  Phys.\ Rev.\ Lett.\  {\bf 102} (2009) 162001
   [Erratum-ibid.\  {\bf 111} (2013) 19,  199905]
  [arXiv:0901.0722 [hep-ph]].

\bibitem{Gardi:2009qi}
  E.~Gardi and L.~Magnea,
  JHEP {\bf 0903} (2009) 079
  [arXiv:0901.1091 [hep-ph]].

\bibitem{Dixon:2009ur}
  L.~J.~Dixon, E.~Gardi and L.~Magnea,
  JHEP {\bf 1002} (2010) 081
  [arXiv:0910.3653 [hep-ph]].

\bibitem{Magnea:2014vha}
  L.~Magnea,
  PoS LL {\bf 2014} (2014) 073
  [arXiv:1408.0682 [hep-ph]].

\bibitem{Gardi:2011yz}
  E.~Gardi, J.~M.~Smillie and C.~D.~White,
  JHEP {\bf 1109} (2011) 114
  [arXiv:1108.1357 [hep-ph]].

\end{thebibliography}
\end{document}